\documentclass[aps,twocolumn,superscriptaddress,floatfix]{revtex4}
\usepackage{amsmath,amssymb}
\usepackage{epsfig}
\usepackage{mathrsfs}

\begin{document}
\title{\boldmath Inverse scattering $J$-matrix approach to
nucleon-nucleus scattering and the shell model}
\author{A. M. Shirokov}
\affiliation{ Skobeltsyn Institute of Nuclear Physics, Moscow State University,
Moscow, 119991, Russia}
\affiliation{Department of Physics and Astronomy,
Iowa State University, Ames IA, 50011-3160, USA}
\author{A. I. Mazur}
\affiliation{
Pacific National University, 136
Tikhookeanskaya street, Khabarovsk 680035, Russia}
\author{J. P. Vary}
\affiliation{Department of Physics and Astronomy,
Iowa State University, Ames IA, 50011-3160, USA}
\author{E. A. Mazur}
\affiliation{
Pacific National University, 136
Tikhookeanskaya street, Khabarovsk 680035, Russia}

  \begin{abstract}
The $J$-matrix inverse scattering approach can be used  as an
   alternative to a
   conventional $R$-matrix in analyzing
   scattering phase shifts and extracting resonance energies and widths
   from experimental data. A great advantage of the $J$-matrix is
   that it provides eigenstates directly related to the ones obtained in
   the shell model in a given model space and with a given value of
   the oscillator spacing $\hbar\Omega$.
   This relationship is of a particular interest
   in the cases when a many-body system does not have a resonant state
   or the resonance is broad and its energy can differ significantly
   from the shell model eigenstate. We discuss the $J$-matrix inverse
   scattering  technique, extend it for the case of charged colliding
   particles and apply it to the analysis of  $n\alpha$ and $p\alpha$
   scattering. The results are compared with  the No-core Shell Model
   calculations of $^5$He and $^5$Li.
\end{abstract}
\maketitle

\section{Introduction}

The $R$-matrix \cite{Lane} is conventionally used in the analysis of
scattering data,
the parameterization of scattering phase shifts and
the extraction of resonant energies and widths from them.
The scattering phase shifts can
also be analyzed in the $J$-matrix formalism of scattering theory \cite{YaFi}.

The inverse scattering oscillator-basis $J$-matrix
approach was suggested in Ref. \cite{Ztmf1}. It was further developed in
Ref. \cite{ISTP} where some useful analytical formulas exploited in this
paper, were derived. The $J$-matrix parameterization of scattering
phase shifts was shown in Ref. \cite{ISTP} to be very accurate in
describing $NN$ scattering data. This  parameterization was used to
construct high-quality non-local $J$-matrix inverse scattering $NN$
potentials JISP6 \cite{JISP6} and JISP16  \cite{JISP16}.

In what follows, we demonstrate that the $J$-matrix can be used for a
high-quality parameterization of scattering phase
shifts in elastic  scattering of nuclear systems  using $n\alpha$ as an
example. The
resonance parameters, its energy and width, can be
easily extracted from the $J$-matrix parameterization.

Resonance energies are conventionally associated with eigenstates above
reaction thresholds obtained in various  nuclear structure models,
e. g., in the shell model. This is well-justified for narrow resonances,
however these eigenstates can
differ significantly from the resonance energies in the case of wide
enough resonances. The $J$-matrix
parameterization naturally provides eigenstates that should be obtained
in the shell model or any other many-body nuclear structure
theory based on the oscillator basis
expansion (e. g., in the resonating group model) to support the experimental
nucleon-nucleus scattering phase shifts in any given model space and
with any given oscillator spacing $\hbar\Omega$.
The shell model eigenstates are provided by the
$J$-matrix phase shift parameterization not only in the case of
resonances, narrow and wide ones,
but also in the case of non-resonant
scattering as well, for example, in the case of $n\alpha$ scattering in the
$\frac12^+$ partial wave.
We will explore these correspondences between the $J$-matrix properties
and results from nuclear structure calculations in some detail below.

Next, we extend the oscillator-basis $J$-matrix inverse scattering approach of
Ref. \cite{ISTP} to the case of charged particles using the formalism
developed in Ref. \cite{Bang}. This extended formalism is shown to work
well in the description of $p\alpha$ scattering and the extraction of $p\alpha$
resonance energies and widths. The shell model eigenstates desired for
the description of the experimental phase shifts, are also provided by the
Coulomb-extended $J$-matrix inverse scattering formalism.

We also carry out No-core Shell Model
\cite{Vary} calculations of $^5$He
and $^5$Li nuclei and compare the obtained eigenstates with the ones
derived from the $J$-matrix parameterizations of $n\alpha$ and $p\alpha$
scattering.

\section{\boldmath $J$-matrix direct and  inverse scattering formalism}

The $J$-matrix formalism \cite{YaFi} utilizes either the oscillator basis or
the so-called Laguerre basis of a Sturmian type. The oscillator basis is
of a particular interest for nuclear applications.
Here we present a sketch of the oscillator-basis $J$-matrix formalism
(more details can be found in Refs. \cite{YaFi,Bang,SmSh}) and some details
of the inverse scattering $J$-matrix approach of Ref. \cite{ISTP}. The
extension of $J$-matrix inverse scattering formalism
to the case of
charged particles is suggested in subsection \ref{ISCoul} while
subsection \ref{ISNCSM} describes how to relate the  $J$-matrix inverse
scattering results to
those of the shell model. 

\subsection{Scattering of uncharged particles }

Scattering in the partial wave with orbital
angular momentum $l$ is governed by a  radial Schr\"odinger equation
 \begin{equation}
  H^l \,u_{l}(E,r)=E\,u_{l}(E,r).
  \label{eq:Sh}
  \end{equation}
Here $r=|{\bf r}|$, $\bf r=r_1-r_2$ is the relative coordinate of
colliding particles and $E$ is the energy of their relative motion.
Within the $J$-matrix formalism, the radial wave function
$u_{l}(E,r)$ is expanded in the oscillator function  series
  \begin{equation}
  u_{l}(E,r) = \sum_{n=0}^{\infty} a_{nl}(E)\, R_{nl}(r),
  \label{eq:row}
  \end{equation}
where the oscillator functions
\begin{multline}
 R_{nl}(r)=(-1)^n\
\sqrt{ \frac{2 n!}{r_0^3
\,\Gamma(n +l+3/2) } }
               \left(\frac{r}{r_0}\right)^{l  
                                         }\\
\times\exp\!{\left(-\frac {r^2}{2r_0^2}\right)}\:
        L_n^{l+\frac12}\!\left(\frac {r^2}{r_0^2}\right),
\label{e24}
\end{multline}
$L^{\alpha}_n(x)$ is the associated Laguerre polynomial, the
oscillator radius  $r_{0}=\sqrt{\hbar / m \Omega}$, and
$m=m_1m_2/(m_1+m_2)$ is the
reduced mass of the
particles with masses $m_1$ and $m_2$.
The wave function in the oscillator representation $a_{nl}(E)$
is a solution of an infinite set of algebraic equations
  \begin{equation}
  \sum_{n'=0}^{\infty}\,(H_{nn'}^{l}-\delta_{nn'}E)\,
  a_{n'l}(E)=0,
  \label{eq:Infsys}
  \end{equation}
where the Hamiltonian matrix elements
$H_{nn'}^l ={T_{nn'}^l+V_{nn'}^{l}}$, the nonzero kinetic energy  matrix elements
\begin{equation}
   \begin{array}{l}
  T_{nn}^l =
    \displaystyle\frac{\displaystyle\hbar\Omega}{\displaystyle 2}\,
                         (2n+l+3/2) , \\
  T_{n+1,n}^l =  T_{n,n+1}^l =
     \displaystyle  -\;\frac{\displaystyle\hbar\Omega}{\displaystyle 2}
     \sqrt{(n+1)(n+l+3/2)} , \end{array}
  \label{Tnm}
  \end{equation}
and the potential energy $V^l$ within the $J$-matrix formalism is
a finite-rank matrix with elements
  \begin{equation}
    \widetilde{V}_{nn'}^l =
\left\{
    \begin{array}{lllcl}
           V_{nn'}^l \ \ \ \ \ &{\rm if} \ \ & n\ &{\rm and}\ & n'\leq\cal N;\\
          \ 0 &{\rm if} & n &{\rm or} & n'>\cal N .
     \end{array} \right.
  \label{trunc}
  \end{equation}


The potential energy matrix truncation (\ref{trunc}) is the only
approximation of the $J$-matrix approach. The kinetic energy matrix is
not truncated, the wave functions are
eigenvectors of the infinite
Hamiltonian matrix $H^l_{nn'}$ which is a superposition of the truncated
potential energy matrix $\widetilde{V}_{nn'}^l$ and the infinite tridiagonal
kinetic energy matrix $T^l_{nn'}$. Note that the Hamiltonian matrix, i. e. both
the kinetic and potential energy matrices, are truncated in conventional
oscillator-basis approaches like the shell model. Hence the $J$-matrix
formalism can be used for a natural extension of the shell model. Note
also that within the inverse scattering $J$-matrix approach,
when the potential energy is represented by the finite matrix (\ref{trunc}), one obtains the exact
scattering solutions, phase shifts and other observables in the continuum spectrum
(see \cite{ISTP} for more details).

The phase shift
$\delta_l$ and the $S$-matrix are expressed in the $J$-matrix formalism as
  \begin{equation}
  \label{osctg}
\tan \delta_l =  - \frac{ S_{{\cal N}l}(E)-{\cal G}_{\cal NN}(E)\,
T^{l}_{{\cal N,\,N}+1}\, S_{{\cal N}+1,\,l}(E) }
  { C_{{\cal N}l}(E)
-{\cal G}_{\cal NN}(E)\, T^{l}_{{\cal N,\,N}+1}\, C_{{\cal N}+1,\,l}(E) } ,
  \end{equation}
\begin{equation}
  S \,=\, \frac
  { C^{(-)}_{{\cal N}l}(E)\,-\,{\cal G}_{\cal NN}(E)\,
T^{l}_{{\cal N,\,N}+1}\, C^{(-)}_{{\cal N}+1,\, l}(E) }
    { C^{(+)}_{{\cal N}l}(E)\,-\,{\cal G}_{\cal NN}(E)\,
T^{l}_{{\cal N,\,N}+1}\, C^{(+)}_{{\cal N}+1,\, l}(E) } ,
  \label{Smat}
  \end{equation}
where
${\cal N}+1$ is the rank of the potential energy matrix (\ref{trunc}), the
kinetic energy matrix elements $T^l_{nn'}$ are given by Eqs.
(\ref{Tnm}), regular $S_{nl}(E)$ and irregular $C_{nl}(E)$
eigenvectors of the infinite kinetic energy matrix are
\begin{equation}
  S_{nl}(E) = 
  \sqrt{ \frac{\pi\, r_{0} \, n!} {\Gamma (n+l+3/2)}}\;
q^{l+1}\, \exp \left(-\frac{q^2}{2}\right) \,
   L_n^{l+\frac12}(q^2),
  \label{eq:Snl}
\end{equation}
\begin{multline}
  C_{nl}(E) = (-1)^l \,
  \sqrt{ \frac{\pi\, r_{0}\,n!}{\Gamma (n+l+3/2)} }\;
  \frac{q^{-l}}{\Gamma (-l+1/2)} \\
\times\exp \left(-\frac{q^2}{2} \right) \Phi (-n-l-1/2,\, -l+1/2;\, q^2),
  \label{eq:Cnl}
\end{multline}
$C^{(\pm)}_{nl}(E) = C_{nl}(E) \pm 
i S_{nl}(E)$,
$\Phi (a,\, b;\, z)$ is a confluent hypergeometric function,
the dimensionless momentum $q=\sqrt{\frac{2E}{\hbar\Omega}}$.
The matrix elements,
  \begin{equation}
  {\cal G}_{nn'}(E) = -\sum_{\lambda =0}^{\cal N}
  \frac{ \langle n|\lambda \rangle
    \langle\lambda  |n'\rangle }
  { E_{\lambda }-E } ,
  \label{oscrm}
  \end{equation}
are expressed through the eigenvalues $E_\lambda$ and eigenvectors
$\langle n|\lambda\rangle$ of the truncated Hamiltonian matrix,
i.~e. $E_\lambda$ and $\langle n|\lambda\rangle$ are obtained by
solving the algebraic problem
\begin{equation}
\sum_{n'=0}^{\cal N} H_{nn'}^l \langle n'|\lambda\rangle = E_\lambda\langle
n|\lambda\rangle, \qquad n\leq {\cal N}.
\label{Alge}
\end{equation}
Only one diagonal matrix element ${\cal G}_{\cal NN}(E)$,
  \begin{equation}
  {\cal G}_{\cal NN}(E) = -\sum_{\lambda =0}^{\cal N}
  \frac{ \langle {\cal N}|\lambda \rangle^2}
  { E_{\lambda }-E } ,
  \label{oscrmNN}
  \end{equation}
is responsible for the phase shifts and the $S$-matrix.

{
The $J$-matrix wave function is given by Eq. (\ref{eq:row})
where 
\begin{equation}
\label{aN1-delta}
a_{nl}(E)=\cos\delta_l\;S_{nl}(E)+\sin\delta_l\;C_{nl}(E) 
\end{equation}
in the `asymptotic region' of the oscillator model space, 
$n\geq \cal N$. Asymptotic behavior \cite{YaFi,Bang,SmSh} of functions 
$\mathscr{S}(E,r)$ and $\mathscr{C}(E,r)$ defined as infinite series,
\begin{multline}
\mathscr{S}(E,r)\equiv\sum_{n=0}^{\infty}S_{nl}(E)\, R_{nl}(r)
=k\, j_l(kr)\\
\mathop{\longrightarrow}\limits_{r\to\infty} \frac{1}{r} \, \sin(kr-\pi l/2)
\label{scrS}
\end{multline}
and
\begin{multline}
\mathscr{C}(E,r)\equiv\sum_{n=0}^{\infty}C_{nl}(E)\, R_{nl}(r)
\mathop{\longrightarrow}\limits_{r\to\infty}
-k\,n_l(kr)\\
\mathop{\longrightarrow}\limits_{r\to\infty}  \frac{1}{r} \, \cos(kr-\pi l/2)
\label{scrC}
\end{multline}
[here $j_l(x)$ and $n_l(x)$ are spherical Bessel and Neumann
functions, and momentum $k=q/r_0$], assures the correct asymptotics of
the wave function (\ref{eq:row}) at positive energies $E$, 
   \begin{multline}
  u_{l}(E,r)\mathop{\longrightarrow}\limits_{r\to\infty} 
k\, [\cos \delta_l\, {j}_{l}(kr) - \sin \delta_l\,{n}_{l}(kr)] \\
\mathop{\longrightarrow}\limits_{r\to\infty}  \frac{1}{r} \, 
               \sin[kr+\delta_l-\pi l/2] .
  \label{asm-gen}
  \end{multline}
In the `interaction
region', $n< \cal N$, $a_{nl}(E)$ are expressed through matrix
elements $ {\cal G}_{n\cal N}(E)$ (see \cite{YaFi,Bang,SmSh} for more details).
However a limited number of
rapidly decreasing with $r$ terms with
$n<{\cal N}$  in expansion (\ref{eq:row})
does not affect asymptotics of the continuum
spectrum wave function.

A similarity between the $J$-matrix and $R$-matrix approaches was
discussed in detail in Ref. \cite{Bang}. Note that the oscillator function
$R_{nl}(r)$ tends to a $\delta$-function in the limit of large 
$n$ \cite{Fil,SmSh},
\begin{gather}
R_{nl}(r) \mathop{\longrightarrow}\limits_{n\to\infty}
\sqrt{2}r_0\, r^{-3/2}\, \delta(r-r_n^{cl}) , 
\label{RtoDelta}
\end{gather}
where 
\begin{gather}
r_n^{cl}=2r_0\sqrt{n+l/2+3/4}
\label{clturp}
\end{gather}
is the classical turning point of the harmonic oscillator eigenstate
described by the function $R_{nl}(r)$. Therefore  expansion
(\ref{eq:row}) describes the wave function
$u_l(E,r)$ at large distances from the origin in a very simple manner:
each term with large enough $n$ gives the amplitude of $u_l(E,r)$ at the 
respective point $r=r^{cl}_n$. Within the $J$-matrix approach, the oscillator
representation  wave functions $a_{nl}(E)$ in the `asymptotic region' of
$n\geq \cal N$ and in the `interaction region' of $n\leq\cal N$ are
matched at $n=\cal N$ \cite{YaFi,Bang,SmSh}. This is equivalent to the
$R$-matrix matching condition 
at the channel radius $r=b$ --- the $J$-matrix formalism reduces to those
of the $R$-matrix with channel radius $b=r^{cl}_{\cal N}$ if $\cal N$ is
asymptotically large. In particular, the function 
${\cal G}_{\cal NN}(E)$ [see (\ref{oscrmNN})] was shown in
Ref. \cite{Bang} to be proportional to the 
$P$-matrix (that is the inverse $R$-matrix) in the limit of
${\cal N}\to\infty$.

At small enough values of $n$, oscillator functions $R_{nl}(r)$
differ essentially from the $\delta$-function. Therefore the $J$-matrix
with realistic values of truncation boundary $\cal N$ differs essentially
from the $R$-matrix approach with realistic channel radius values
$b$. It appears that the $J$-matrix formalism with its matching
condition in the oscillator model space, is somewhat better suited to 
traditional nuclear structure models like the shell model.
}

In the inverse scattering $J$-matrix approach,  the phase
shifts $\delta_l$ are supposed to be known at any energy $E$ and we are
parameterizing them by Eqs. (\ref{osctg}), (\ref{eq:Snl}),
(\ref{eq:Cnl}), and (\ref{oscrmNN}), i. e. one should find the eigenvalues
$E_\lambda$ and the eigenvector components $\langle {\cal N}|\lambda\rangle$
providing a good
description of the phase shifts. If the set of
$E_\lambda$ and $\langle {\cal N}|\lambda\rangle$ values is known,
i. e. the function ${\cal G}_{\cal NN}(E)$ is completely defined,
the $S$-matrix poles are obtained by  solving numerically an obvious
equation,
  \begin{equation}
   C^{(+)}_{{\cal N}l}(E)\,-\,{\cal G}_{\cal NN}(E)\,
T^{l}_{{\cal N,\,N}+1}\, C^{(+)}_{{\cal N}+1,\, l}(E) = 0  ,
  \label{Spole}
  \end{equation}
where solutions for $q $ 
(or $E=\frac12 q^2\hbar\Omega$)  should be searched for in the desired
domain of the complex plane.

Knowing the phase shifts $\delta_l$ in a large enough energy interval
$0\leq E<E_{\max}$, one gets the set of eigenenergies $E_\lambda$,
$\lambda=0,$ 1, ...~, $\cal N$ by solving numerically the equation
\begin{equation}
 a_{{\cal N}+1,l}(E)=0,
 \label{Elam}
\end{equation}
where $a_{{\cal N}+1,l}(E)$ is given by Eq. (\ref{aN1-delta}).
The equation (\ref{Elam}) has exactly ${\cal N}+1$ solutions.
The last components $\langle{\cal  N}|\lambda\rangle$ of the eigenvectors
$\langle n|\lambda\rangle$  responsible for the phase shifts and the
$S$-matrix, are obtained as
\begin{equation}
|\langle{\cal  N}|\lambda\rangle|^2=
\frac{a_{{\cal N}l}(E_\lambda)}{\alpha_l^{\lambda}\;T_{{\cal N,\,N}+1}^l},
 \label{Nlambda}
\end{equation}
where
\begin{equation}
 \alpha^{\lambda}_l=
\left.\frac{d\,a_{{\cal N}+1,\,l}(E)}{d\,E}\right|_{E=E_\lambda}.
\label{allambda}
\end{equation}

The physical meaning of the Eqs. (\ref{Elam}), (\ref{Nlambda}) is
the following. The equation (\ref{Elam})
guarantees that the phase
shifts $\delta_l$ exactly reproduce the experimental phase
shifts at the energies $E=E_{\lambda}$. The equation  (\ref{Nlambda})
fixes the derivatives of the phase shifts  $\frac{d\delta_l}{dE}$ at
the energies $E=E_{\lambda}$ fitting them
exactly to the derivatives of the experimental phase shifts at the same
energies.

The solutions $E_{\lambda}$ and $\langle{\cal  N}|\lambda\rangle$,
$\lambda=0,$ 1, ...~, $\cal N$ depend strongly on the values of the
oscillator spacing $\hbar\Omega$ and $\cal N$, the size of the inverse
scattering potential matrix. Larger
values of $\cal N$ and/or  $\hbar\Omega$,
imply a larger
energy interval $0\leq E<E_{\max}$ where the phase shifts
are reproduced by the $J$-matrix parameterization
(\ref{osctg}).

A Hermitian Hamiltonian generates a set of normalized eigenvectors
$\langle n|\lambda\rangle$ fitting the completeness relation,
\begin{equation}
\sum_{\lambda =0}^{\cal N}|\langle{\cal N}|\lambda\rangle|^2=1.
\label{Complete}
\end{equation}
Experimental phase shifts generate a set of $\langle{\cal  N}|\lambda\rangle$,
$\lambda=0,$ 1, ...~, $\cal N$ that usually does not fit
Eq. (\ref{Complete}).
It is likely that the interval of energy values
used to find the sets of $E_{\lambda}$ and $\langle{\cal  N}|\lambda\rangle$,
spreads beyond
the thresholds where new channels are opened.
Thus inelastic channels
are present in the system
suggesting the Hamiltonian should become non-Hermitian.
The approach proposed
in Ref. \cite{ISTP}, suggests to fit Eq. (\ref{Complete}) by changing
the value of the component
$\langle{\cal  N}|\lambda=\cal N\rangle$ corresponding to
the largest among the energies $E_{\lambda}$ with $\lambda=\cal N$. This
energy $E_{\lambda=\cal N}$ is usually larger than $E_{\max}$, the
maximal energy in the interval $0\leq E<E_{\max}$
where the  experimental phase shifts are available. Therefore changing
$\langle{\cal  N}|\lambda=\cal N\rangle$ should not spoil 
the phase shift description in the desired interval
of energies below $E_{\max}$;
more over, one can also vary
subsequently the energy  $E_{\lambda=\cal N}$ to improve the description
of the phase shifts in the interval $0\leq E<E_{\max}$.

We are not discussing here the construction of the inverse scattering
potential
but point the interested reader to Ref. \cite{ISTP}.
{
We note only that if the  
construction of the $J$-matrix inverse scattering potential is desired,
one should definitely fit Eq. (\ref{Complete}), otherwise the
construction of the Hermitian interaction is impossible. In our
applications to $n\alpha$ and $p\alpha$ scattering we are interested
only in the $J$-matrix parameterization of scattering phase shifts;
hence we can avoid renormalization of the component
$\langle{\cal  N}|{\lambda=\cal N}\rangle$. Nevertheless, we found out
that this renormalization improves the phase shifts description at
energies $E$ not close to $E_\lambda$ values. All the results presented
below were obtained with the help of  Eq. (\ref{Complete}).
}

\subsection{Charged particle scattering }\label{ISCoul}


In the case of
a charged  projectile scattered by a charged target, the interaction
between them
is a superposition of a short-range nuclear interaction,
$V^{Nucl}$, and the Coulomb interaction, $V^C$:
\begin{equation}
V=V^{Nucl}+V^C.
\label{VnuclC}
\end{equation}
The Coulomb interaction between proton and nucleus is conventionally
described as (see, e. g., \cite{Kuku})
\begin{equation}
V^{C}=Z e^2\,
\frac{\mathrm{erf}(r/x_0)}{r} .
 \label{VC1}
\end{equation}
In the case of $p\alpha$
scattering discussed below, $Z=2$ and $x_0=1.64$~fm \cite{Kuku}.

{
The long-range Coulomb interaction (\ref{VC1}) requires
some modification of the  oscillator-basis $J$-matrix formalism
described in the previous subsection. In the case of charged particle
scattering, the wave function $u_{l} (E,r)$ at asymptotically large
distances takes a form:}
  \begin{equation}
   u_{l} (E,r)=  k\,[ \cos \delta_l\, {f}_{l}(\zeta,kr)
          - \sin \delta_l\, {g}_{l}(\zeta,kr)] ,
  \label{eq:ascoul}
  \end{equation}
where
  \begin{gather}
  {f}_{l}(\zeta,kr)=\frac{1}{kr}\:
F_{l}(\zeta,kr),\\
  {g}_{l}(\zeta,kr)= -\:\frac{1}{kr}\:
G_{l}(\zeta,kr),
 \end{gather}
$F_{l}(\zeta,kr)$ and $G_{l}(\zeta,kr)$
are regular and irregular Coulomb functions respectively,
{
and Sommerfeld parameter $\zeta = Ze^{2} m /k$. Instead  of functions 
$\mathscr{S}(E,r)$ and $\mathscr{C}(E,r)$, one can introduce functions
$\mathscr{F}(E,\zeta,r)$ and $\mathscr{G}(E,\zeta,r)$
 defining them as infinite series,
\begin{gather}
\mathscr{F}(E,\zeta,r)\equiv\sum_{n=0}^{\infty}F_{nl}(E,\zeta)\, R_{nl}(r)
=k\, f_l(\zeta,kr)
\label{scrF}
\end{gather}
and
\begin{gather}
\mathscr{G}(E,\zeta,r)\equiv\sum_{n=0}^{\infty}G_{nl}(E,\zeta)\, R_{nl}(r)
\mathop{\longrightarrow}\limits_{r\to\infty} -k\,g_l(\zeta,kr),
\label{scrG}
\end{gather}
in order to use $F_{nl}(E,\zeta)$ and $G_{nl}(E,\zeta)$ in constructing
continuum spectrum wave functions by means of Eq. (\ref{eq:row}). Such
an approach was proposed by the Kiev group in Ref. \cite{Okhr}. Within this
approach, the $J$-matrix matching condition at $n=\cal N$ becomes
much more complicated, resulting in difficulties in designing an inverse
scattering approach and in shell model applications. In practical
calculations, the approach of Ref. \cite{Okhr} requires the use of much
larger values of $\cal N$, i. e. a huge extension of the model space when
solving the algebraic problem (\ref{Alge}), that makes it incompatible
with the shell model applications. Therefore it is desirable to find
another way to extend our approach on the case of charged particle scattering.
}

We use here the formalism of Ref. \cite{Bang} to allow for the Coulomb
interaction in the oscillator-basis $J$-matrix theory.
{
The idea of the approach is very simple. Suppose there are
a long-range $V$ and a short-range $V^{Sh}$ potentials that are
indistinguishable at distances $0<r<b$. In this case, the potential
$V^{Sh}$ generates a wave function fitting exactly (up to an overall
normalization factor) that of the long-range potential $V$ at $r<b$. If
the only difference between $V$ and  $V^{Sh}$ at distances $r>b$ is the
Coulomb interaction, then one can equate the logarithmic derivatives of
their wave functions at $r=b$ and use the resulting equation to express
the long-range potential phase shifts $\delta_l$ in terms of the short-range
potential phase shifts $\delta_l^{Sh}$ or vice versa. Note that the
phase shifts $\delta_l^{Sh}$ can be obtained within the standard $J$-matrix
approach discussed in the previous subsection. The recalculation of the
phase shifts $\delta_l^{Sh}$ into $\delta_l$ (or vice versa) appears to be the
only essential addition in
formulating such a direct (or inverse) Coulomb-extended $J$-matrix formalism.

To implement this idea,}
we introduce a channel radius $b$ large enough to neglect the nuclear
interaction $V^{Nucl}$ at distances  $r \geq b$, i. e. $b\geq R_{Nucl}$,
where $R_{Nucl}$ is the range of the potential  $V^{Nucl}$. In the
asymptotic region $r \geq b$, the radial wave function $u_{l} (E,r)$
is given by Eq. (\ref{eq:ascoul}).

At short distances $r\leq b$, the wave function  $u_{l} (E,r)$ coincides
with $u^{Sh}_{l}(E,r)$, the one generated by the auxiliary potential
  \begin{equation}
  V^{Sh} = \left\{ \begin{array}{cl}
  V = V^{Nucl} + V^{C},  & \mbox{ $r \leq b$} \\
  0,                   & \mbox{ $r > b$}
  \end{array} \right. ;
  \ \   b \geq R_{Nucl}
  \label{eq:pot1}
  \end{equation}
obtained by truncating the Coulomb potential $V^C$ at $r=b$. The wave
function  $u^{Sh}_{l}(E,r)$ behaves asymptotically as a wave function
obtained with a short-range interaction,
   \begin{equation}
  u^{Sh}_{l}(E,r) = k[\cos \delta^{Sh}_l\, {j}_{l}(kr)  -
           \sin \delta^{Sh}_l\, {n}_{l}(kr)] , \quad  b \geq R_{Nucl}.
  \label{eq:asm}
  \end{equation}
The $J$-matrix formalism described in the previous
subsection, 
should be used to calculate the function  $u^{Sh}_{l}(E,r)$,
the  auxiliary phase shift $\delta^{Sh}_l$ and the respective
auxiliary $S$-matrix $S^{Sh}$.

Matching the functions  $u_{l} (E,r)$ and $u^{Sh}_{l}(E,r)$ at $r=b$,
the phase shift $\delta_l$ can be expressed through $\delta^{Sh}_l$
\cite{Bang}:
  \begin{equation}
  \tan \delta_l = \frac{ W_{b}({j}_{l},{f}_{l})-
   W_{b}({n}_{l},{f}_{l})\tan \delta^{Sh}_l }
 {W_{b}({j}_{l},{g}_{l})-W_{b}({n}_{l},{g}_{l})
 \tan \delta^{Sh}_l } ,
  \label{tantan}
  \end{equation}
where quasi-Wronskian
\begin{multline}
\label{Wronsk}
 W_{b}({j}_{l},{f}_{l})
\equiv \left. \left\{
\frac{d}{dr}[{j}_{l}(kr)]\,{f}_{l}(\zeta,kr)\right.\right.\\
\left.\left.-{j}_{l}(kr)\,\frac{d}{dr}{f}_{l}(\zeta,kr)\right\}
\right|_{r=b},
\end{multline}
and $W_{b}({n}_{l},{f}_{l})$,
$W_{b}({j}_{l},{g}_{l})$ and $W_{b}({n}_{l},{g}_{l})$
are expressed similarly. The $S$-matrix is given by
  \begin{equation}
  S = \frac{ W_{b}(h^-_{l},g^-_{l})-
   W_{b}(h^+_{l},g^-_{l})\, S^{Sh} }
 {W_{b}(h^-_{l},g^+_{l})-W_{b}(h^+_{l},g^+_{l})\,
 S^{Sh} } ,
  \label{Smatcl}
  \end{equation}
where $h^{\pm}_{l}(kr)=-n_{l}(kr)\pm ij_{l}(kr)$,
$g^{\pm}_{l}(\zeta,kr)=-g_{l}(\zeta,kr)\pm if_{l}(\zeta,kr)$, and
the quasi-Wronskians
$W_{b}(h^{\pm}_{l},g^{\pm}_{l})$ are defined by analogy with
Eq. (\ref{Wronsk}). The $S$-matrix poles are obtained by solving
the equation
  \begin{equation}
  \label{Spolecl}
W_{b}(h^-_{l},g^+_{l})-W_{b}(h^+_{l},g^+_{l})\,
 S^{Sh} = 0
  \end{equation}
in the complex energy plane.

This formalism involves a free parameter, the channel radius $b$, used
for construction of the auxiliary potential $V^{Sh}$.
As mentioned above, $b$  should be
taken larger than the range of the short-range nuclear interaction
$V^{Nucl}$. On the other hand, the truncated
$({\cal N}+1) \times ({\cal N}+1)$ Hamiltonian matrix
$H_{nn'}^l$ ($n,n'=0$, 1, ...~, $\cal N$) used to calculate  the sets of
eigenvalues $E_{\lambda}$ and
eigenvectors $\langle{\cal N}|\lambda \rangle$ by
solving the algebraic problem (\ref{Alge}),
should carry information about the jump of potential $V^{Sh}$ at the
point $r=b$. Therefore $b$ should be chosen less than approximately
$r_{\cal N}$, the classical turning
point of the oscillator function $R_{{\cal N}l}(r)$,
the function with the
largest range in the set of oscillator
functions $R_{nl}(r)$, $n=0$, 1, ...~, $\cal N$ used for the
construction of the truncated Hamiltonian matrix
${H}_{nn'}^l$~$(n,n'\leq\cal N)$. In
a practical calculation, one should
study convergence with a set of $b$ values and pick up the $b$ value
providing the most stable and best-converged results. As
shown in
Ref. \cite{Bang}, the phase shift $\delta_l$ calculated at some energy
$E$ as a function of channel radius $b$, usually has a plateau in the interval
$R_{Nucl}<b<r_{\cal N}^{cl}$ that reproduces well the exact values of $\delta_l$.

In the inverse scattering approach, first, we fix a value of the channel
radius $b$ and transform experimental phase shifts $\delta_l$ into the
set of auxiliary phase shifts $\delta^{Sh}_l$:
  \begin{equation}
  \tan \delta^{Sh}_l = \frac{ W_{b}({j}_{l},{f}_{l})-
   W_{b}({j}_{l},{g}_{l})\tan \delta_l }
 {W_{b}({n}_{l},{f}_{l})-W_{b}({n}_{l},{g}_{l})
 \tan \delta_l }.
  \label{tanret}
  \end{equation}
Equation (\ref{tanret}) can be easily obtained by inverting Eq.
(\ref{tantan}). Next, we employ the inverse scattering approach of the
previous subsection to calculate the sets of $E_{\lambda}$ and
$\langle{\cal  N}|\lambda\rangle$ using auxiliary phase shifts
$\delta^{Sh}_l$ as an input. The $J$-matrix parameterization of the
phase shifts $\delta_l$ is given by Eq.  (\ref{tantan}), the $S$-matrix
poles can be calculated through Eq. (\ref{Spolecl}).

\subsection{\boldmath$J$-matrix and the shell model}\label{ISNCSM}

Up to
this point we have been discussing the $J$-matrix formalism supposing the
colliding particles to be structureless. In applications to the $n\alpha$
and $p\alpha$ scattering and relating the respective $J$-matrix inverse
scattering results to the shell model, we should have in mind that the $\alpha$
particle consists of 4 nucleons identical to the scattered nucleon and
the five-nucleon wave function should be antisymmetrized. The $J$-matrix
solutions and the expressions (\ref{osctg}) for the phase shifts and
(\ref{Smat}) for the
$S$-matrix [or expressions (\ref{tantan}) and
(\ref{Smatcl}) in the case when both the projectile and the target are
charged], can be used in the case of
scattering of complex systems comprising identical fermions. The
components $\langle {\cal N}|\lambda \rangle$ entering expression
(\ref{oscrmNN}) for the function ${\cal G}_{\cal NN}(E)$ become, of
course, much more complicated: they  now appear to be some particular
components of the many-body eigenvector. However, we are not interested
here in the microscopic many-body structure of the components
$\langle{\cal N}|\lambda\rangle$;
we shall obtain them by fitting the
$n\alpha$ and $p\alpha$ phase shifts in the $J$-matrix inverse
scattering approach.

We focus our attention here on  other important
ingredients entering expression
(\ref{oscrmNN}) for ${\cal G}_{\cal NN}(E)$, the eigenenergies
$E_\lambda$,
related to the energies of the states in the
combined many-body system, i.~e. in the $^5$He or $^5$Li nucleus in the
case of $n\alpha$ or $p\alpha$ scattering respectively, obtained in
the shell model or any other many-body approach utilizing the oscillator
basis. One should have however in mind that $E_\lambda$ entering
Eq. (\ref{oscrmNN})  correspond to the kinetic energy of relative motion,
i.~e. they are always
positive, while many-body microscopic approaches generate eigenstates
with absolute energies, e.~g. all the states in $^5$He and $^5$Li with
excitation energies below approximately 28 MeV (the $\alpha$-particle
binding energy) will be generated negative.
Therefore, before comparing with the
set of $E_\lambda$ values, one  should perform a simple recalculation of
the shell model eigenenergies by adding to them  the $^4$He binding energy;
or alternatively one can use the set of
$E_\lambda$ values to calculate the respective set of energies defined
according to the shell model definitions by subtracting the $^4$He
binding energy from
each of $E_\lambda$. The physical meaning of
transforming these to the shell model
scale of values for $E_\lambda$ is
to provide the values required from shell model calculations
in order to reproduce the desired phase shifts.

The comparison of the inverse scattering $J$-matrix analysis with the
shell model results is useful, of course, only  if
the same $\hbar\Omega$ value is used both in the  $J$-matrix and in
the shell model and model spaces of these approaches are properly correlated. A
traditional notation for the model space within the
shell model is $N_{\max}\hbar\Omega$ where $N_{\max}$ is the excitation
oscillator quanta.
In the case of the $J$-matrix, we use, also
traditionally, $\cal N$, the principal quantum number of the highest
oscillator function $R_{{\cal N}l}(r)$ included in the `interaction
region' of the oscillator model space
where the potential energy matrix
elements are retained.
The following
expressions relate $N_{\max}$ and $\cal N$
in the cases of  $\frac32^-$
and $\frac12^-$ partial waves ($p$ waves) and  $\frac12^+$ partial wave
($s$ wave):
\begin{gather}
N_{\max}=2{\cal N},\quad {\cal N}=0,1,...\,, \quad\text{$\frac32^-$
and $\frac12^-$ partial waves,}\\
N_{\max}=2{\cal N}-1,\quad {\cal N}=1,2,...\,, \quad \frac12^+\text{
 partial wave. }
\end{gather}

{
Below we are using shell model type $N_{\max}\hbar\Omega$ notations
for labeling both $J$-matrix and shell model results.
}

\section{\boldmath Analysis of $n\alpha$ scattering phase shifts}
\subsection{\boldmath $\frac32^-$ phase shifts}

We start discussion of our $J$-matrix analysis of $n\alpha$
scattering
with the $\frac32^-$ phase shifts.

The $J$-matrix inverse scattering approach was well-tested in the
nucleon-nucleon
($NN$) scattering in Ref. \cite{ISTP}. In the case of $NN$-scattering,
the phase shifts are well established in a wide range of energies up to
350 MeV in laboratory system. The goal of Ref. \cite{ISTP} was to fit
scattering phase shift in the entire interval of energies
$0\leq E_{lab}\leq 350$ MeV using
the smallest possible potential matrices
or, equivalently,
the smallest possible values of $N_{\max}$ (or $\cal N$). In the case of
nucleon-$\alpha$ scattering, the phase shifts are known in a small
energy interval up to $E_{lab}=20$
MeV and in some cases up to 25 MeV. On
the other hand, to compare the $J$-matrix analysis with the shell model
results, we are interested in large enough values of $N_{\max}$ and in
$\hbar\Omega$ values reasonable for shell model applications. As a
result, we face a problem of insufficient data: some solutions
$E_\lambda$ of Eq. (\ref{Elam}) should be
allowed far outside the interval
of known phase shifts $\delta_l$.
The required phase shifts should be known together with their
derivatives at the energies around $E=E_\lambda$ just to find these
solutions $E_\lambda$ [see Eq. (\ref{aN1-delta})] and respective
eigenvector components $\langle{\cal  N}|\lambda\rangle$ [see Eqs.
(\ref{Nlambda}) and (\ref{allambda})].

\begin{figure}
\centerline{\epsfig{file=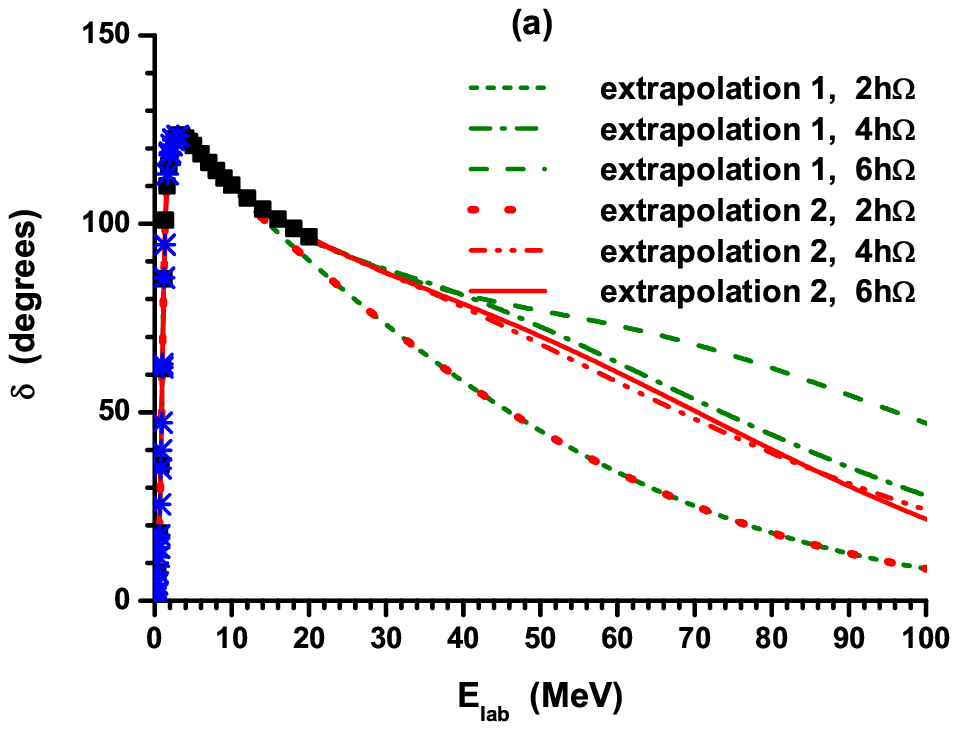,width=0.47\textwidth}}
\centerline{\epsfig{file=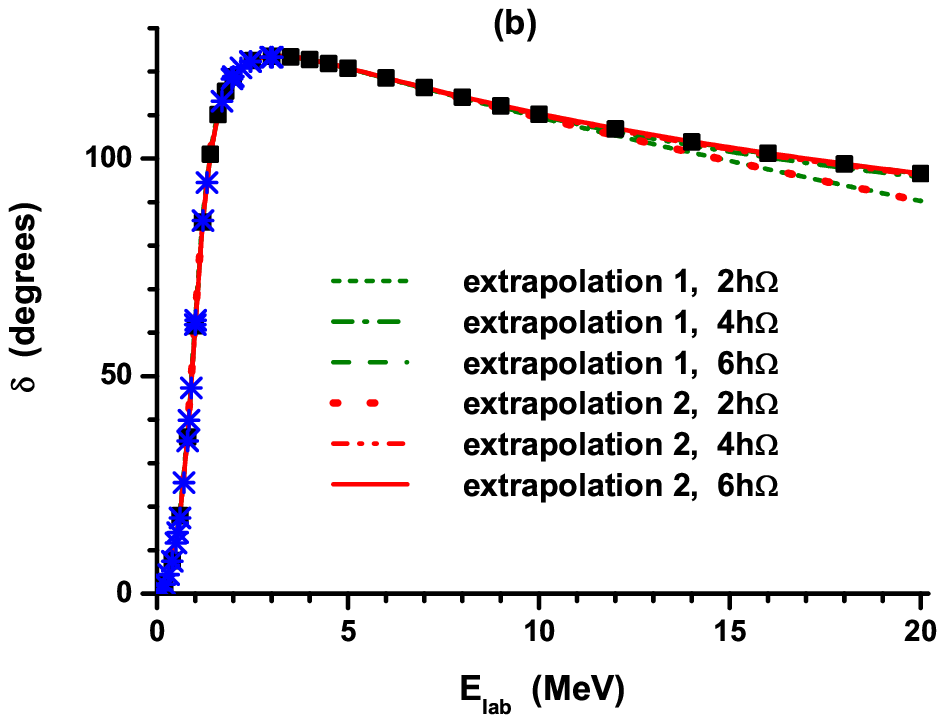,width=0.47\textwidth}}\vspace{-1ex}
\caption{(Color online) Experimental $\frac32^-$ $n\alpha$ phase shifts
 from Refs. \cite{ArndtDLRop} (stars) and \cite{NPA287}
 (filled squares) and $J$-matrix parameterizations with two arbitrary
 extrapolations of phase shifts for $E_{lab}>20$ MeV obtained with
$\hbar\Omega=20$ MeV in  $2\hbar\Omega$,  $4\hbar\Omega$
 and  $6\hbar\Omega$  model spaces. Both
  panels present the same results
 in different energy scales.}
\label{na-ext}
\end{figure}

We address the
problem of data insufficiency by
an extrapolation of the data
outside the energy
interval of known phase shifts.
The $J$-matrix parameterizations presented in
Fig. \ref{na-ext} were obtained with $\hbar\Omega=20$ MeV in various
model spaces.
In each case two different extrapolations were used for the
phase shifts at energies $E_{lab}>20$ MeV, however the experimental
phase shifts below $E_{lab}=20$ MeV are equivalently well described if
$N_{\max}$ is large enough.
The deviation of the parameterization
from the experiment is seen at energies $E_{lab}>10$ MeV only in the
case of the $2\hbar\Omega$ model
space, the smallest among  all model spaces
presented in Fig. \ref{na-ext}, and even in this case the deviation is
small enough.
This is not surprising since the phase shifts given by
Eqs. (\ref{osctg}) and (\ref{oscrmNN}) in the low-energy interval are
governed mostly by the $E_\lambda$ values from the same interval and by the
respective eigenvector components $\langle{\cal N}|\lambda\rangle$.
These $E_\lambda$ and $\langle{\cal N}|\lambda\rangle$ values are
determined by Eqs. (\ref{Elam}) and (\ref{Nlambda}) locally, i.~e. they
are independent from the phase shift extrapolation. Note that in the
case of the $2\hbar\Omega$ model space, both $E_\lambda$ values lie in
the energy interval of known phases, hence the parameterization in this
model space is  completely independent from the
extrapolation and  the two parameterizations
obtained in this model space  coincide.

The resonance energy
$E_{res}$ and width $\Gamma$ calculated by locating the $S$-matrix pole
by solving Eq. (\ref{Spole}),
are seen from Table \ref{tab-extr} to be very stable and
insensitive to the extrapolation of the phase shifts.


\begin{table}
\caption{The energy $E_{res}$ and width $\Gamma$ (both in MeV) of the
 $\frac32^-$ resonance in the $n\alpha$ scattering obtained with
 $\hbar\Omega=20$ MeV  in various model spaces $N_{\max}\hbar\Omega$
 with two different extrapolations of the phase shifts.}
\begin{ruledtabular}
\begin{tabular}{ccccc}
 &\multicolumn{2}{c}{Extrapolation 1}
              &\multicolumn{2}{c}{Extrapolation 2} \\ \cline{2-5}
 $N_{\max}$   & $E_{res}$  & $\Gamma$    & $E_{res}$  & $\Gamma$     \\ \hline
  6 &  0.7713  &   0.6437     &    0.7718  &  0.6435 \\
  8 &  0.7719  &   0.6451     &    0.7715  &  0.6454 \\
 10 &  0.7707  &   0.6417      &   0.7708  &  0.6416 \\ 
\end{tabular}
\label{tab-extr}
\end{ruledtabular}
\end{table}

\begin{figure}\vspace{-1ex}
\centerline{\epsfig{file=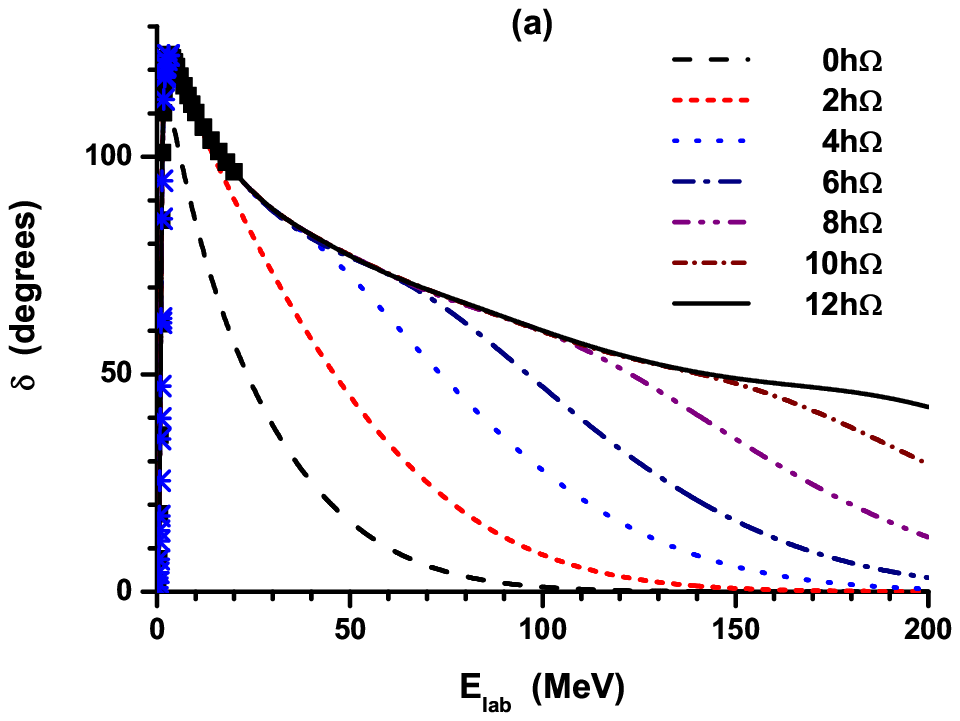,width=0.46\textwidth}}
\centerline{\epsfig{file=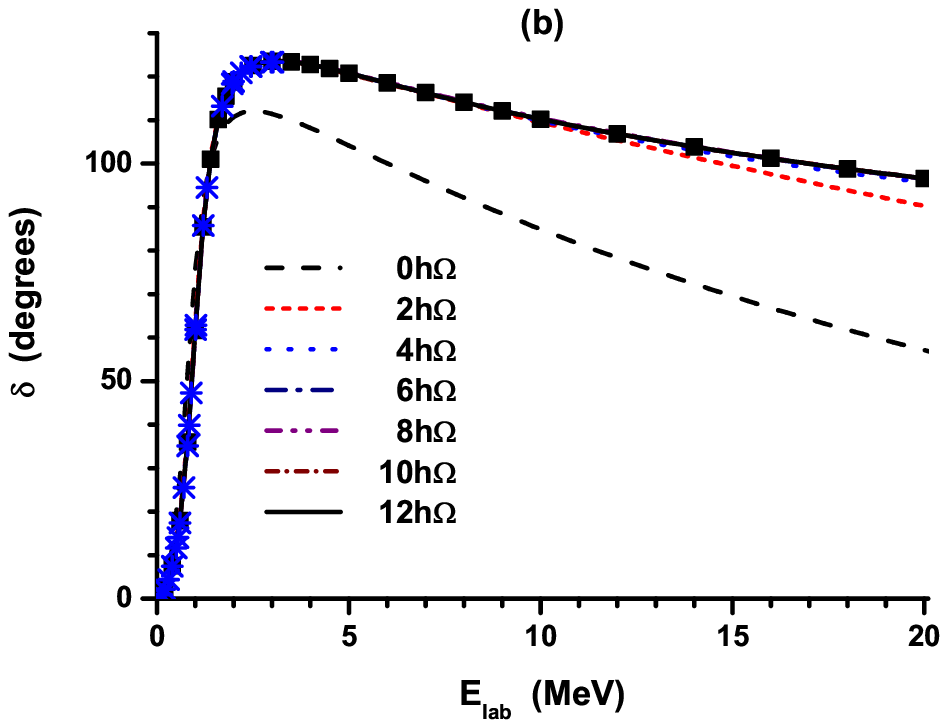,width=0.46\textwidth}}
\centerline{\epsfig{file=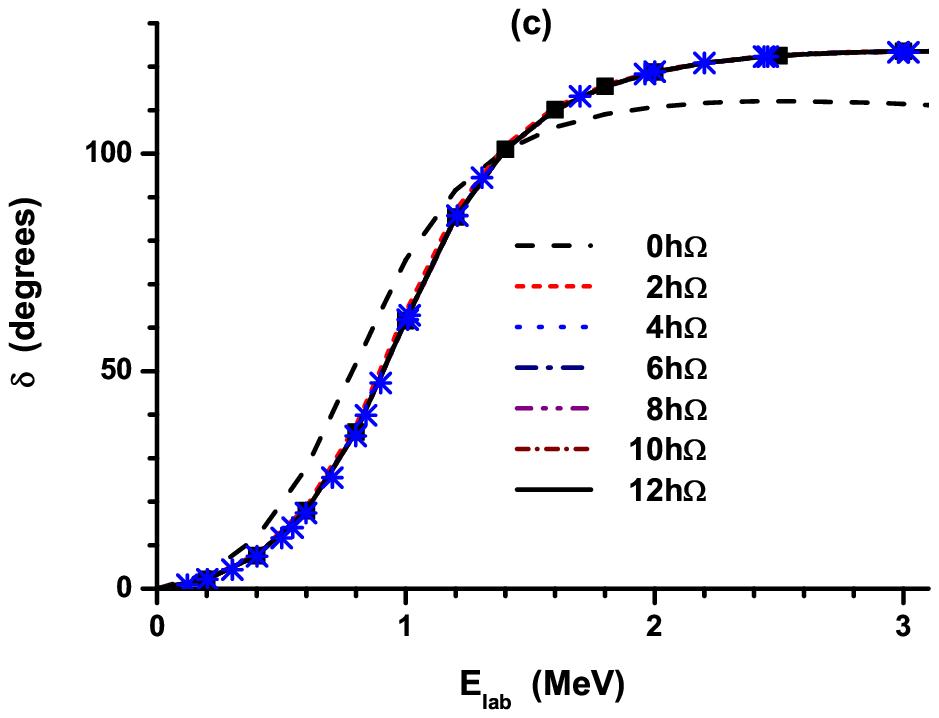,width=0.46\textwidth}}\vspace{-1.5ex}
\caption{(Color online) The $J$-matrix parameterization of the
 $\frac32^-$ $n\alpha$ phase shifts obtained with $\hbar\Omega=20$ MeV
 in various model spaces. Different panels present the same results in
 different energy scales. Experimental phase shifts: stars~---
 Ref. \cite{ArndtDLRop},  filled squares~---
Ref. \cite{NPA287}.}\vspace{-2ex}
\label{phnap3h20}
\end{figure}

The same insensitivity of the phase description in the desired energy
interval to the phase shift extension outside this interval,
is also inherent for other $n\alpha$ partial waves.
Hence, we shell not waste space by discussing this issue in the respective 
subsections below.
{
We should probably just note here that the need for data extrapolation
arose only due to our desire to compare the $J$-matrix results with the
shell model ones; it is this desire that pushes us to use large enough
model spaces and 
$\hbar\Omega$ values. If one is interested only in getting a high
quality $J$-matrix parameterization of the phase shifts and in extracting
resonance parameters, smaller model spaces and/or smaller $\hbar\Omega$
values can be used without a loss of accuracy and without a need to have
phase shifts outside the experimentally known energy interval.
}

A resulting practical approach is to  extrapolate the phase shifts in any
reasonable way outside the energy interval  where they are known
in order to obtain the $J$-matrix
parameterization of the phase
shifts within this interval of known phases and to derive  resonance
parameters in the same energy interval. Using such extrapolation, we
study a dependence of the  $J$-matrix phase
shift parameterization on the size of the model space. As is seen from
Fig. \ref{phnap3h20}, larger model spaces make it possible to describe
the extrapolated phase shifts up to larger energies. The experimental
data are perfectly reproduced in $4\hbar\Omega$ and larger model
spaces. We fail to reproduce the experiment for $E_{lab}>12$ MeV in the
$2\hbar\Omega$ model space. Note however that deviations from the
experiment are not very large and we obtain a very good
description of the phase shifts at laboratory energies below 12 MeV
including the resonance region. The smallest possible $0\hbar\Omega$
model space fails to provide a reasonable description of the phase
shifts at all energies.

\begin{figure*}
{\epsfig{file=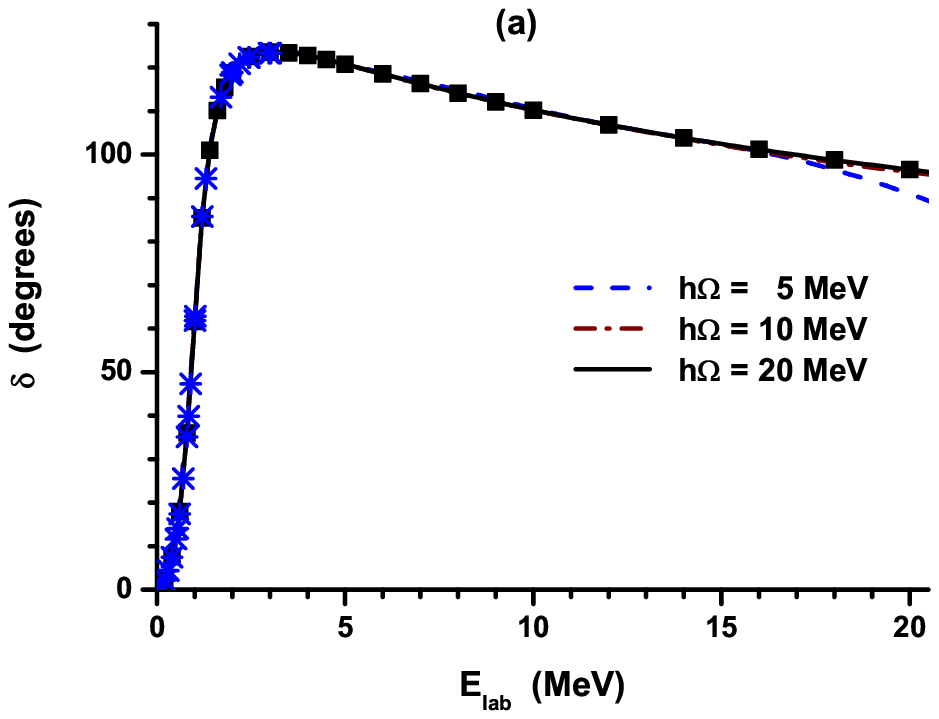,width=0.47\textwidth}}\hfill
{\epsfig{file=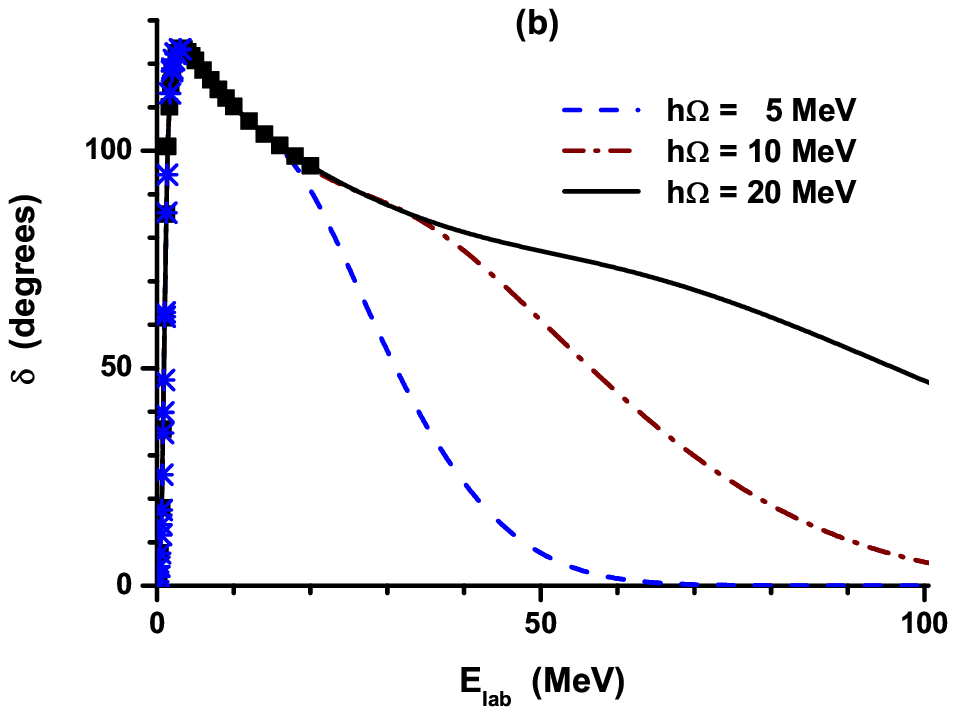,width=0.47\textwidth}}
\caption{(Color online) The $J$-matrix parameterization of the
 $\frac32^-$ $n\alpha$ phase shifts obtained in the $6\hbar\Omega$ model
 space with different values of oscillator spacing $\hbar\Omega$.
See Fig. \ref{phnap3h20} for details.}
\label{phnap3n4}
\end{figure*}

As mentioned above, the description of the phase shifts can be
extended to larger energies not only by using larger model spaces but
also by using larger
$\hbar\Omega$ values. This is illustrated by Fig. \ref{phnap3n4}. Even
with $\hbar\Omega=5$ MeV we manage to describe the phase shifts in the
$6\hbar\Omega$ model space up to approximately $E_{lab}=17$ MeV. The
description of all experimentally known phase shifts is perfect in this
model space with $\hbar\Omega=10$ MeV  and larger.

\begin{figure*}
{\epsfig{file=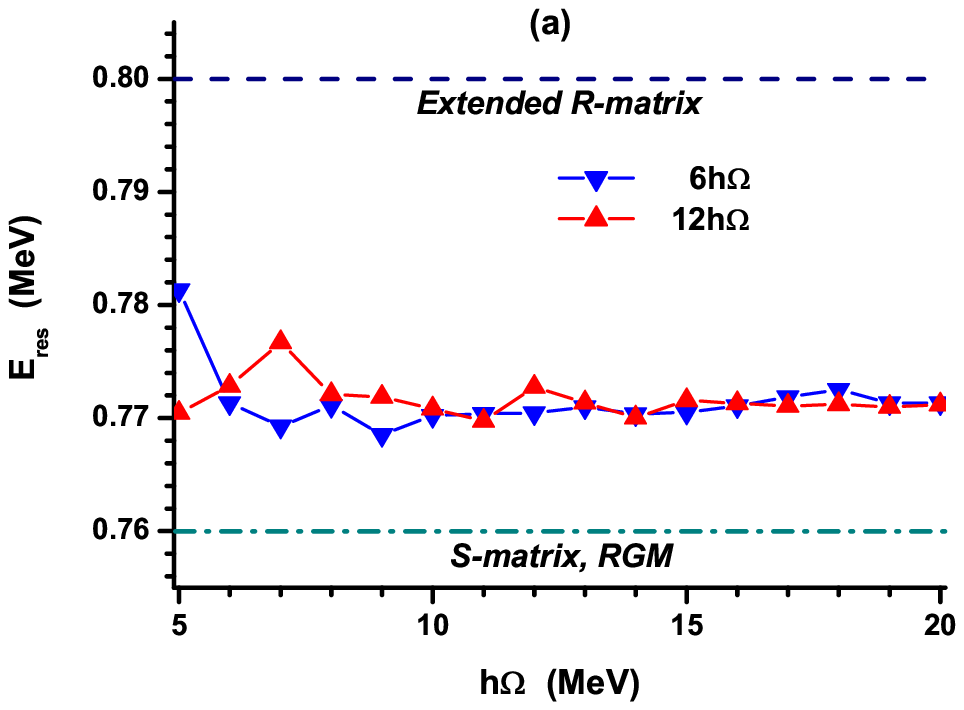,width=0.47\textwidth}}\hfill
{\epsfig{file=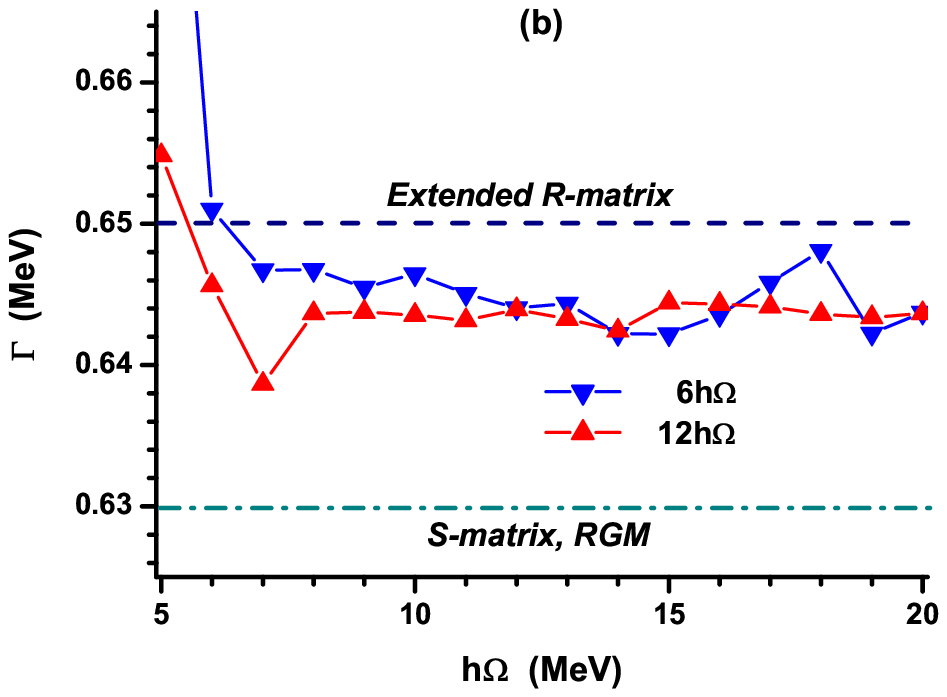,width=0.47\textwidth}}{
\epsfig{file=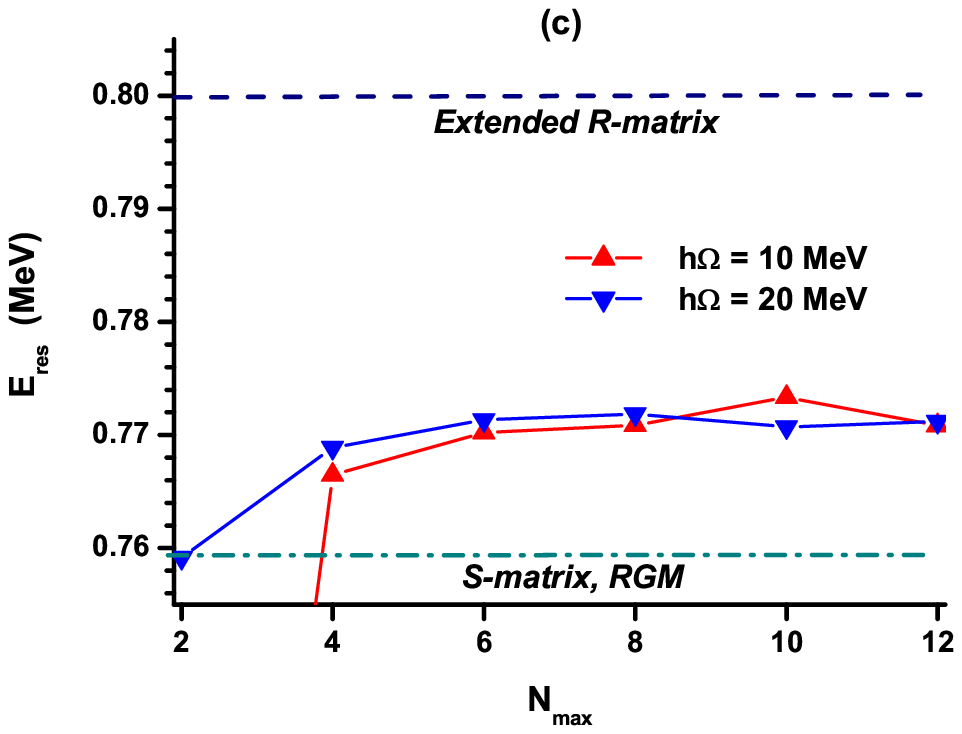,width=0.47\textwidth}}\hfill
{\epsfig{file=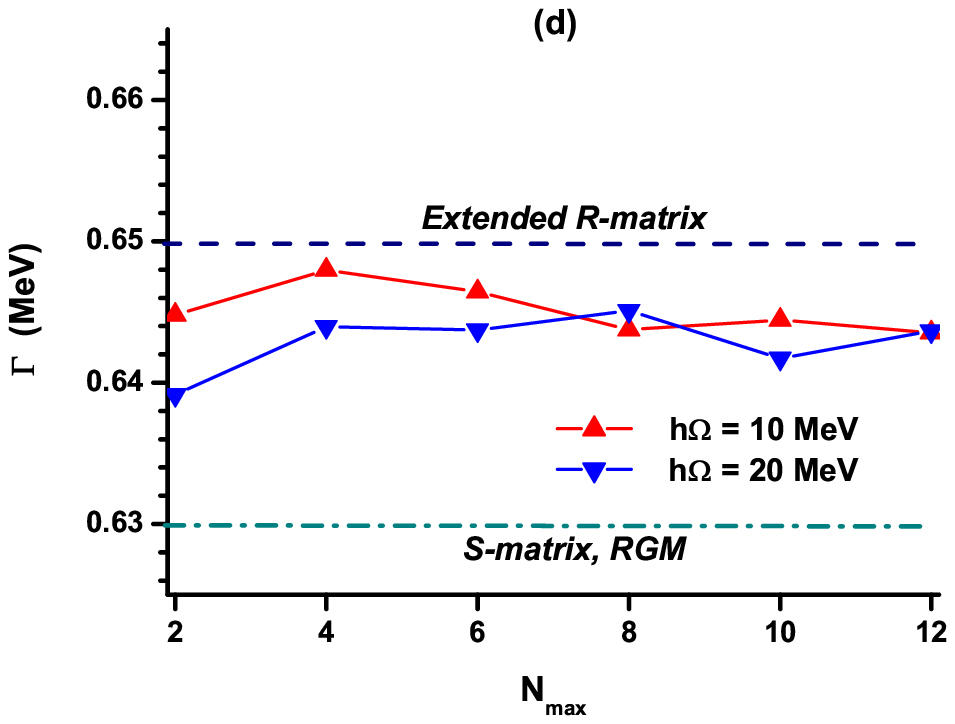,width=0.47\textwidth}}
\caption{(Color online) The  $n\alpha$ $\frac32^-$ resonance energy in
 the center-of-mass frame
 (left) and width (right) obtained by calculating the position of the
 $S$-matrix pole by means of the $J$-matrix parameterizations with
 different $\hbar\Omega$ values (upper panels) and in different model spaces
(lower panels). Horizontal lines present the results of Ref. \cite{Rmatr}:
the analysis of the resonance parameters in the extended
 $R$-matrix approach (dashed) and calculations of the $S$-matrix pole
 position in the Resonating Group Method (dash-dotted).}
\label{ergamp3}
\end{figure*}

The results of calculations of the $S$-matrix  pole position are
presented in Fig. \ref{ergamp3}. The calculated resonance energy $E_{res}$ and
width $\Gamma$ are seen to be very stable in a wide range of
$\hbar\Omega$ values and model spaces (note a very detailed energy scale
in Fig. \ref{ergamp3}). Our results are in a very good correspondence with the
results of a detailed study of Ref. \cite{Rmatr}. The authors of this
paper performed
Resonating Group Method calculations of $N\alpha$ scattering with
phenomenological Minnesota $NN$ interaction fitted to reproduce with
high precision the $n\alpha$ and $p\alpha$ phase shifts and calculated
the position of the $S$-matrix pole.
The extended multichannel
$R$-matrix analysis of $^5$He and $^5$Li including two-body channels
$N+\alpha$ and $d+t$ or $d+\rm^3He$ along with pseudo-two-body
configurations to represent the breakup channels $n+p+t$ or
$n+p+\rm^3He$, was also performed in Ref. \cite{Rmatr} using data of
various authors on the differential elastic scattering cross sections,
polarization, analyzing-power and polarization-transfer measurements
together with neutron total cross sections. Our very simple $J$-matrix
analysis utilizing only the elastic scattering phase shifts, is
competitive in quality of resonance parameter description with these
extended studies of Ref. \cite{Rmatr}.

\begin{figure}\vspace{-1ex}
\centerline{\epsfig{file=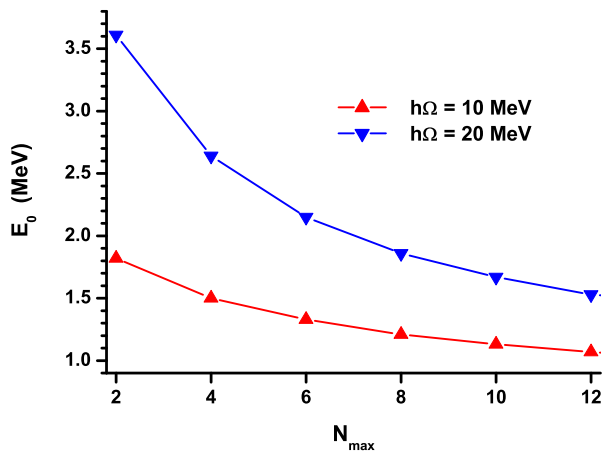,width=0.47\textwidth}}\vspace{-2ex}
\caption{(Color online) The lowest state $E_{\lambda=0}$ obtained in the
$J$-matrix parameterization of the
 $\frac32^-$ $n\alpha$ phase shifts with different $\hbar\Omega$  values
 in various model spaces.}
\label{Elam0}
\end{figure}

We note that while the phase shifts and resonance parameters are very stable,
the energies $E_\lambda$ entering
Eq. (\ref{oscrmNN}) vary essentially with $\hbar\Omega$ and model
space. In particular, this is true  for  the lowest of these energies
$E_{\lambda=0}$ shown in Fig. \ref{Elam0} (note a very large difference
in energy scales in Figs. \ref{ergamp3} and  \ref{Elam0}).
This energy being obtained in shell model studies,
would be
associated traditionally with the resonance energy
$E_{res}$. Such a conventional association is clearly incorrect: this
lowest eigenstate $E_{\lambda=0}$ differs significantly in energy from
$E_{res}$ while the phase shifts and resonance energy and width are well
reproduced;  just  this energy
$E_{\lambda=0}$, very different from $E_{res}$, is needed to have a
perfect description of scattering data and resonance parameters including
$E_{res}$ itself.
The $E_{\lambda=0}$ dependencies of the type shown in Fig. \ref{Elam0}
are inherent in other partial waves and in the case of $p\alpha$ scattering.
We study the  $E_{\lambda=0}$ dependencies on
$\hbar\Omega$ and model space in more detail
below in Section \ref{SM-El} where we compare them with the results of
No-core Shell Model  calculations.

\subsection{\boldmath $\frac12^-$ phase shifts}

\begin{figure}\vspace{-2ex}
\centerline{\epsfig{file=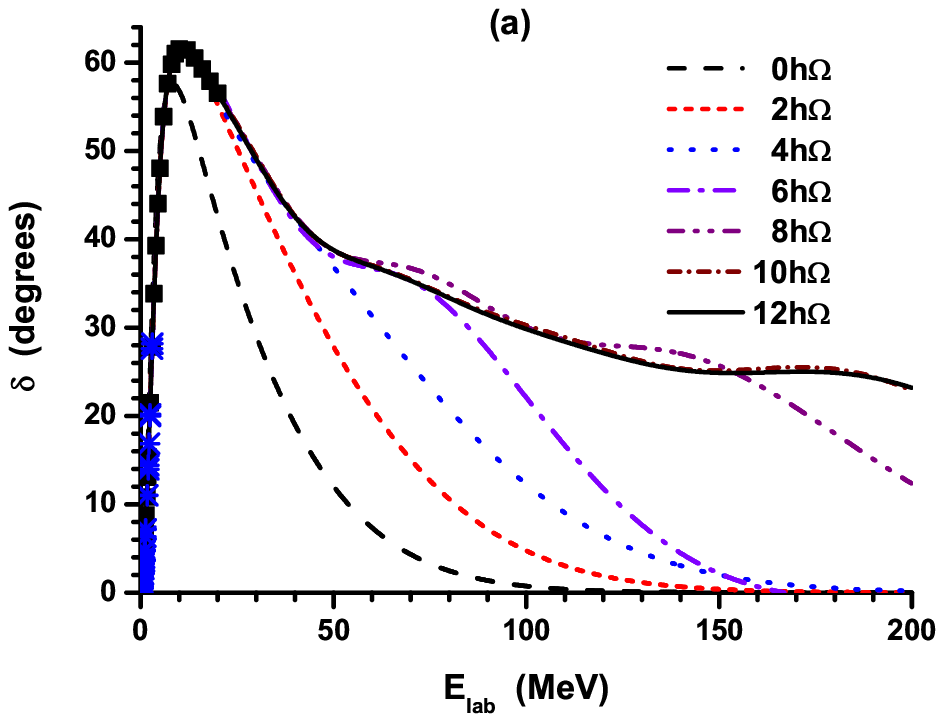,width=0.47\textwidth}}
\centerline{\epsfig{file=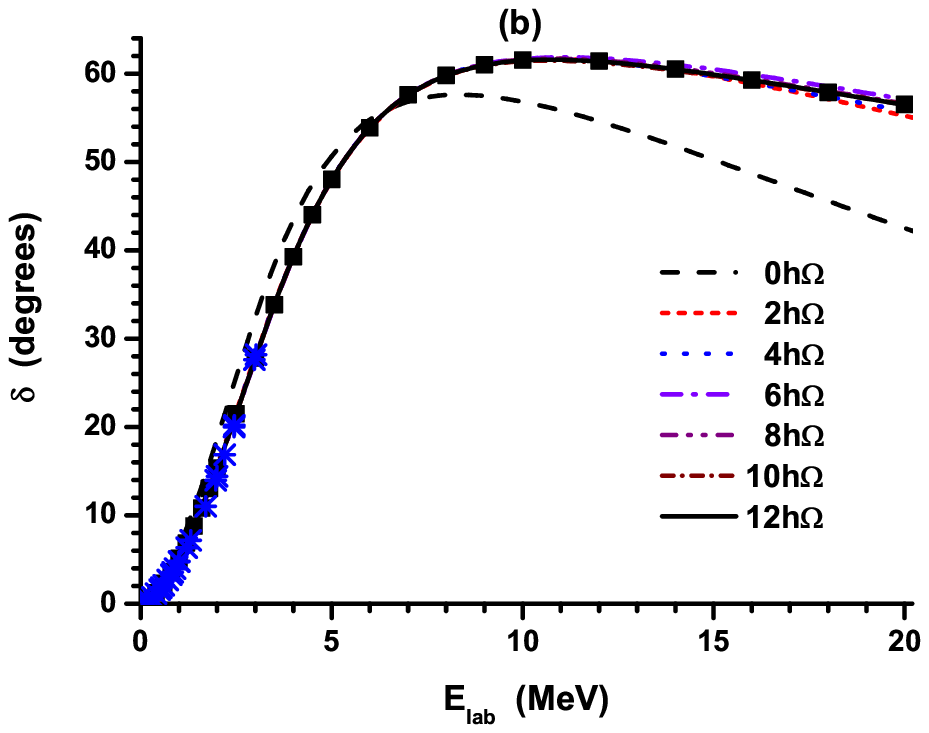,width=0.47\textwidth}}
\centerline{\epsfig{file=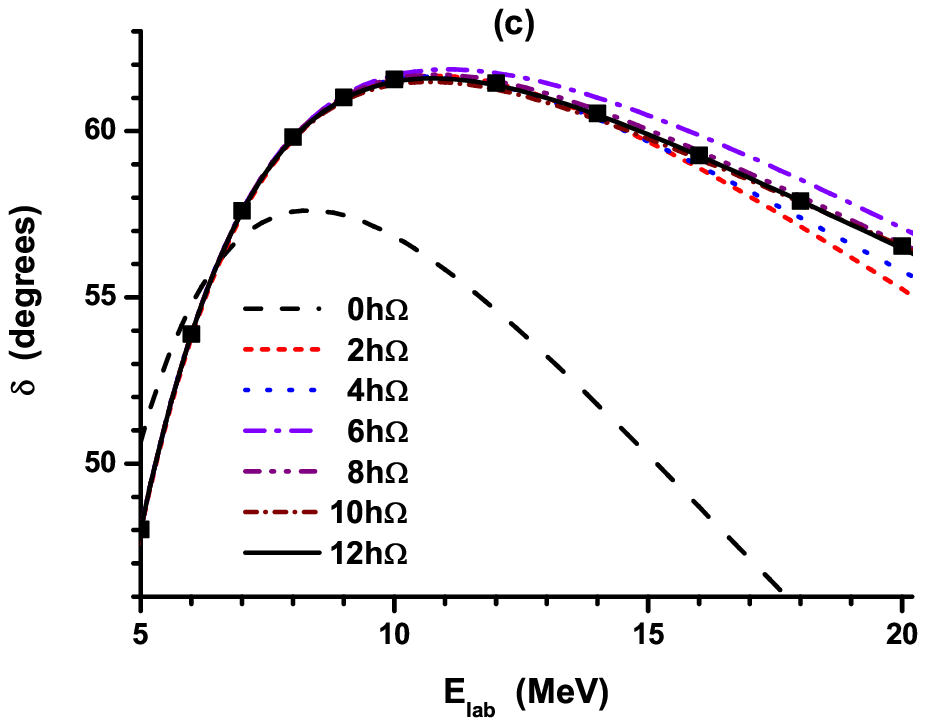,width=0.47\textwidth}}\vspace{-1.5ex}
\caption{(Color online) The $J$-matrix parameterization of the
 $\frac12^-$ $n\alpha$ phase shifts obtained with $\hbar\Omega=20$ MeV
 in various model spaces. See Fig. \ref{phnap3h20} for details.}\vspace{-1ex}
\label{phnap1h20}
\end{figure}

\begin{figure*}
{\epsfig{file=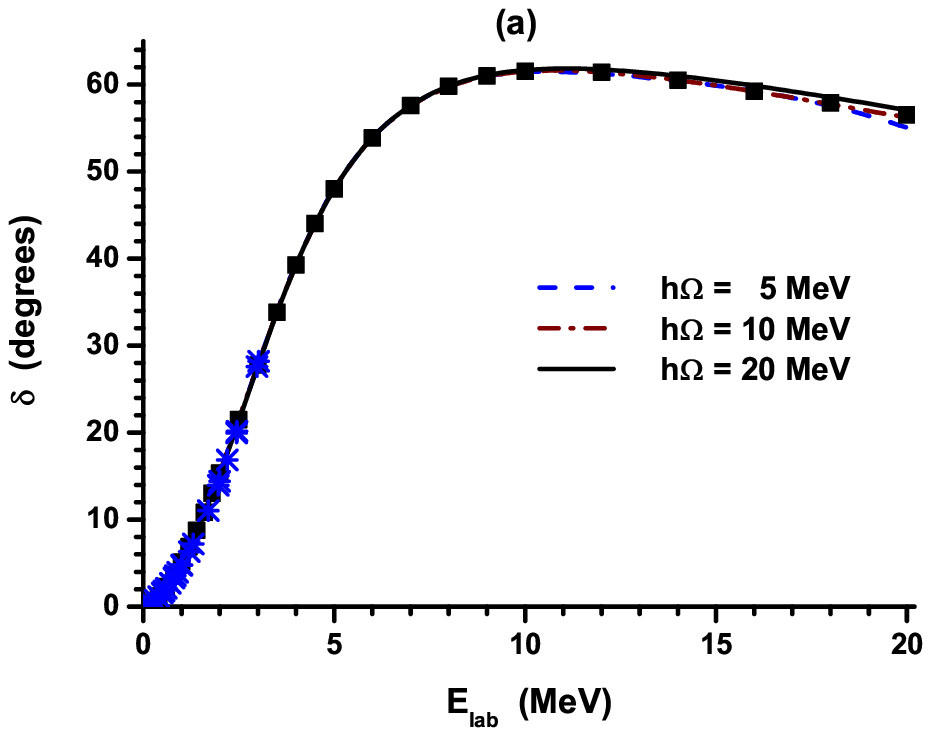,width=0.47\textwidth}}\hfill
{\epsfig{file=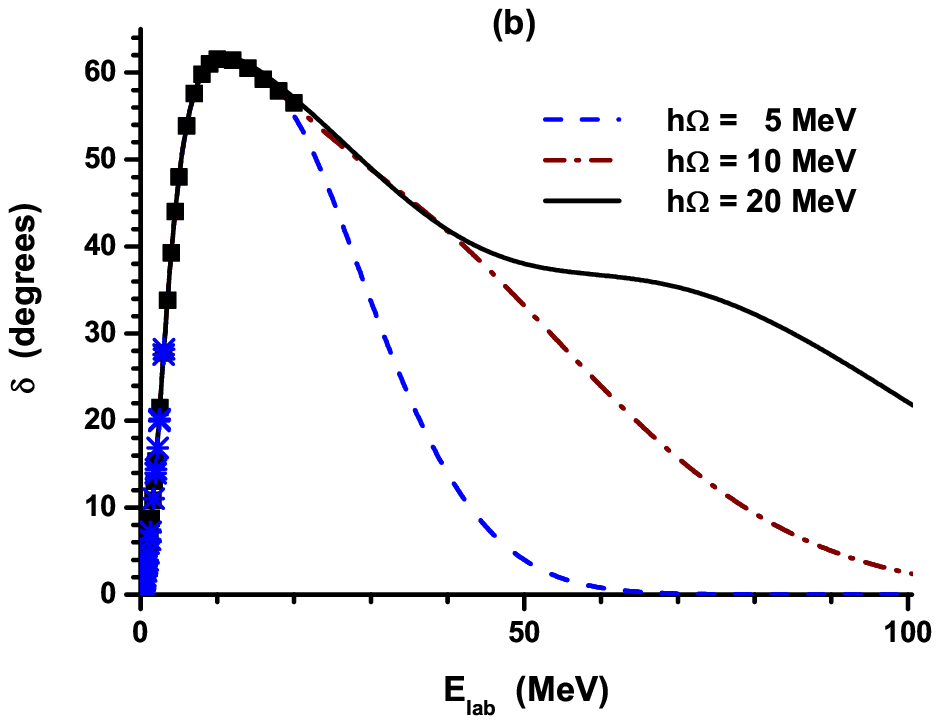,width=0.47\textwidth}}
\caption{(Color online) The $J$-matrix parameterization of the
 $\frac12^-$ $n\alpha$ phase shifts obtained in the $6\hbar\Omega$ model
 space with different values of oscillator spacing $\hbar\Omega$. See
 Fig. \ref{phnap3n4} for details.}
\label{phnap1n4}
\end{figure*}

\begin{figure*}
{\epsfig{file=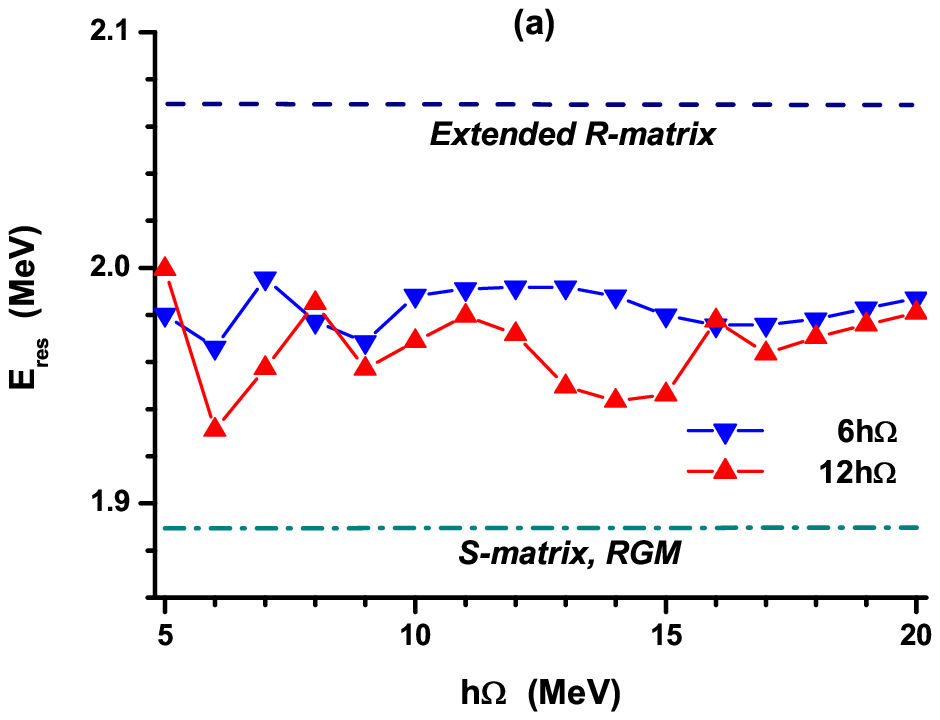,width=0.47\textwidth}}\hfill
{\epsfig{file=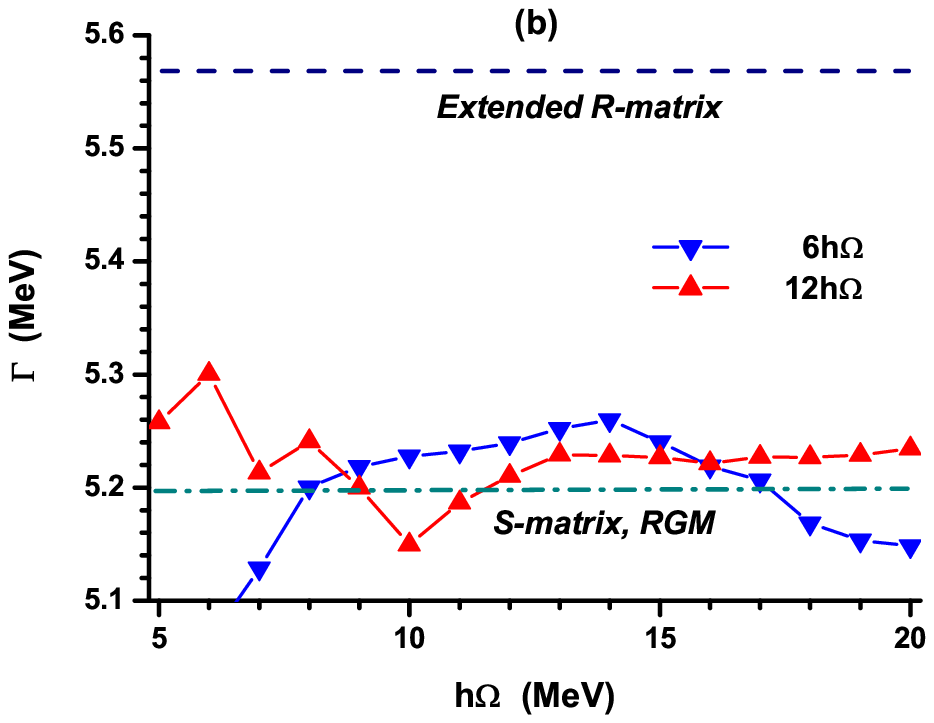,width=0.47\textwidth}}
{\epsfig{file=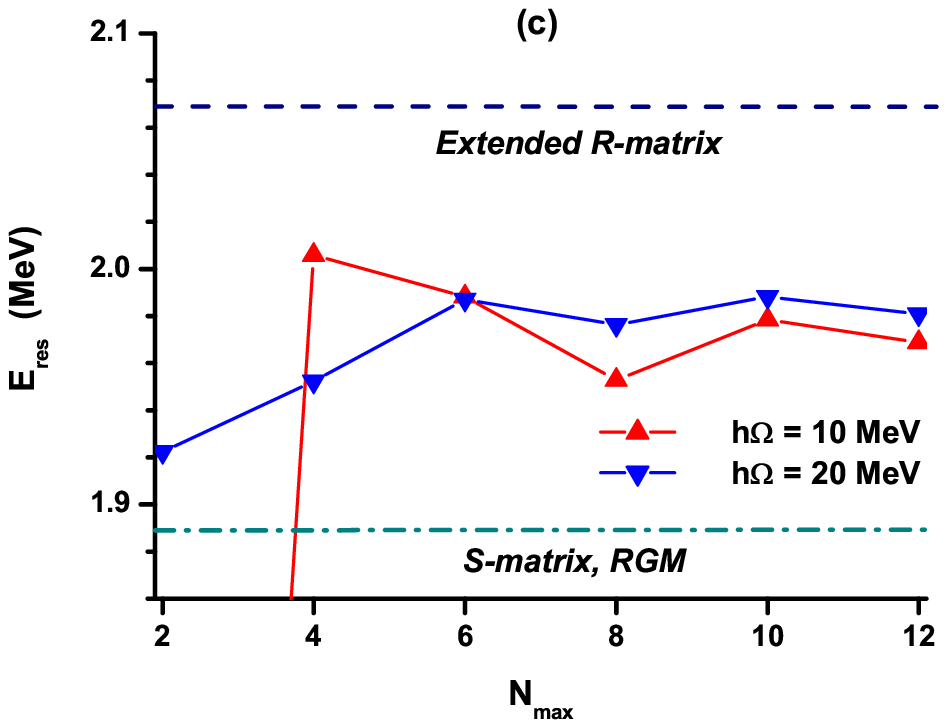,width=0.47\textwidth}}\hfill
{\epsfig{file=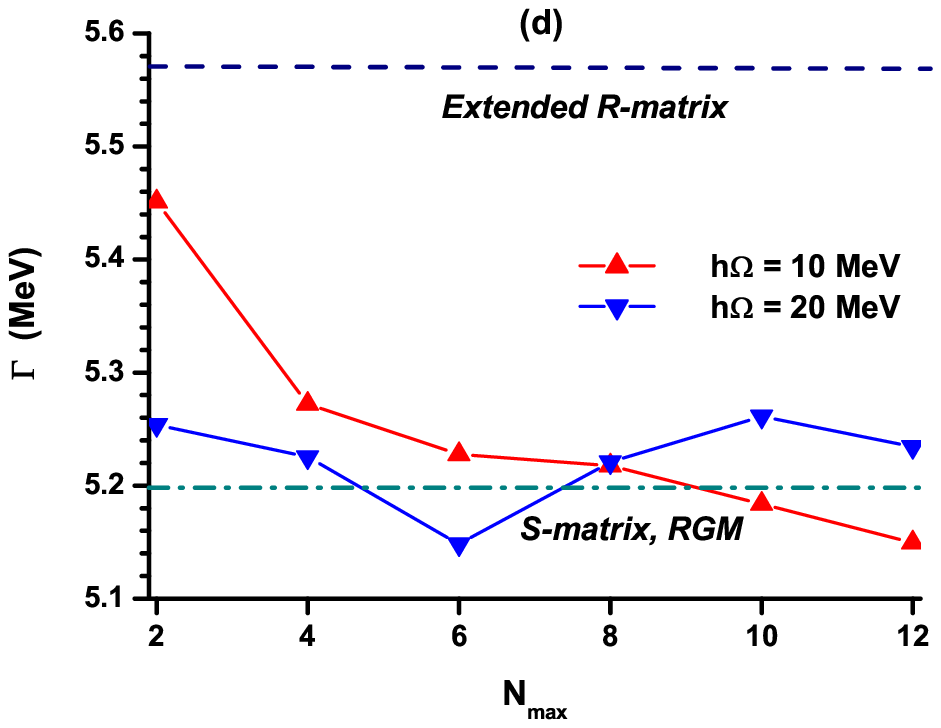,width=0.47\textwidth}}
\caption{(Color online) The  $n\alpha$ $\frac12^-$ resonance energy in
 the center-of-mass frame
 (left) and width (right) obtained by calculating the position of the
 $S$-matrix pole by means of the $J$-matrix parameterizations with
 different $\hbar\Omega$ values (upper panels) and in different model spaces
(lower panels). See Fig. \ref{ergamp3} for details.}
\label{ergamp1}
\end{figure*}

We present in Figs. \ref{phnap1h20} and \ref{phnap1n4} respectively the
$J$-matrix parameterizations
of $n\alpha$ $\frac12^-$ phase shifts obtained with the same
$\hbar\Omega$ in different model spaces and with different $\hbar\Omega$
values in the same model space. The description of the  $\frac12^-$
phase shifts with different
$\hbar\Omega$ values and in different model spaces follows the same
patterns as in the case of the $\frac32^-$ phase shifts. The only
difference is  that a high-quality description of the phase
shifts  at energies $E_{lab}>10$ MeV is attained in  larger model
spaces. However, in the $8\hbar\Omega$ and larger model spaces the
description of all known phase shifts is perfect.

Figure \ref{ergamp1} presents the results of our calculations of the
$\frac12^-$ resonance energy and width.  The variations
of $E_{res}$ and $\Gamma$ with increasing
$\hbar\Omega$ or model
space are larger than in the case of the $\frac32^-$ resonance; note
however that the energy of the $\frac12^-$ resonance and its width are also
much larger.
At any rate, the variations of resonance parameters are not large
and our results for $E_{res}$ and $\Gamma$ are stable enough with
respect to the choice of $\hbar\Omega$ value and model space. The
energy and width of the $\frac12^-$ resonance
also compare well with the results of Ref. \cite{Rmatr}.

\subsection{\boldmath $\frac12^+$ phase shifts}

In
describing the $\frac12^+$ phase shifts, one should have in mind
that the lowest
$s$ states are occupied in the
$\alpha$-particle and due
to the Pauli principle these states should be inaccessible to the
scattered nucleon. There are two conventional approaches to the problem of
the Pauli forbidden  $s$ state in the $n+\alpha$ system.
The first approach is to add a phenomenological repulsive
term to the $s$ wave component of the $n\alpha$ potential (see,
e. g., \cite{Rep-na}). This
phenomenological repulsion excludes the Pauli forbidden state in
the $n+\alpha$ system and is supposed to simulate the
Pauli principle effects in more complicated cluster systems.
Another approach is to use
deep attractive $n\alpha$ potentials
that support the Pauli forbidden $s$ state in the $n+\alpha$
system (see \cite{Forb-na,Kuku,Forb-na2}). In the cluster model studies,
the Pauli forbidden state is excluded by projecting it out
\cite{Kuku,Forb-na2,LurAnn}.

In our $J$-matrix inverse scattering approach, we can simulate both the
potentials with repulsive core and with a forbidden state. In the first
case, when the system does not have a bound state, we go on with the
same procedure as in the above cases of $\frac32^-$ and $\frac12^-$
partial waves; the energy dependence of the input $\frac12^+$ phase
shifts is responsible for generating proper details of the $n\alpha$
interaction potential matrix. In the other case, the simplest way to
simulate the presence of the forbidden state in the system is to suppose
that this state is described by a pure $0s_{1/2}$ oscillator wave
function. The energy of the forbidden state is equal in this case to the
Hamiltonian matrix element $H_{00}^{l=0}$ which is of no interest for us
in this study, all the matrix elements
$H_{0n}^{l=0}$ and $H_{n0}^{l=0}$ should be set equal to zero to guarantee
the orthogonality of the forbidden state to scattering states which  have
the wave functions given by the expansion (\ref{eq:row}) where the
$0s_{1/2}$ oscillator state is missing, i. e.  $a_{n=0,\,l=0}(E)=0$ for all
energies $E>0$. Within this model, the forbidden state \cite{IS} does not
contribute to the function ${\cal G}_{\cal NN}(E)$ [see
Eq. (\ref{oscrmNN})] since the component
$\langle {\cal N}|\lambda=0 \rangle=0$. In the inverse scattering
approach, we use the first $\cal N$ solutions of Eq. (\ref{Elam})
disregarding the highest in energy solution $E_{{\cal N}+1}$ while
constructing the function ${\cal G}_{\cal NN}(E)$.

\begin{figure}
\centerline{\epsfig{file=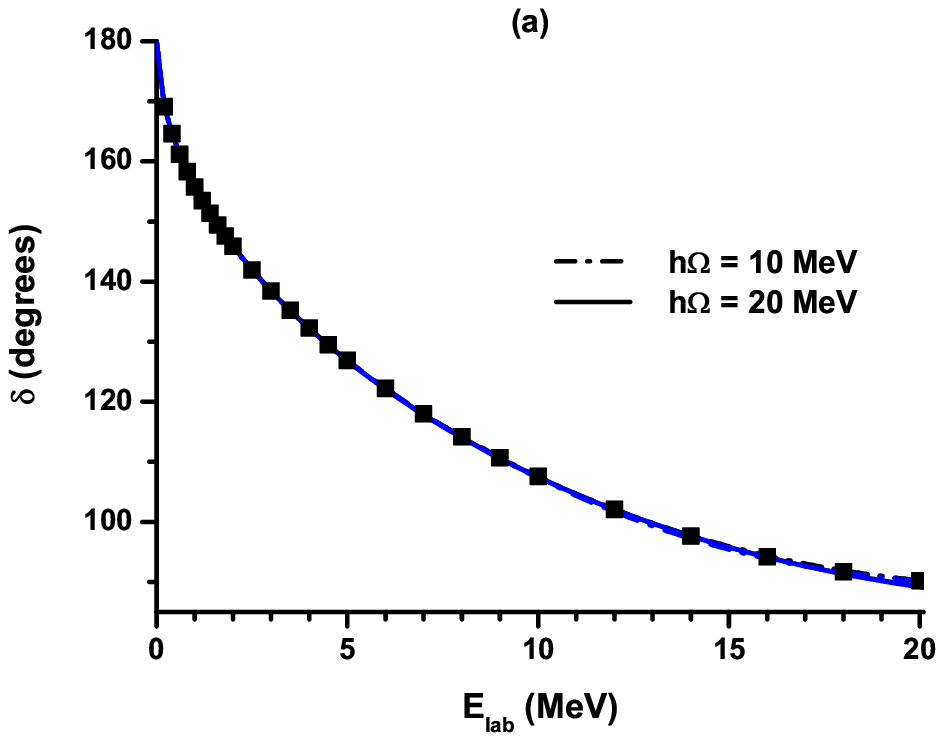,width=0.47\textwidth}}
\centerline{\epsfig{file=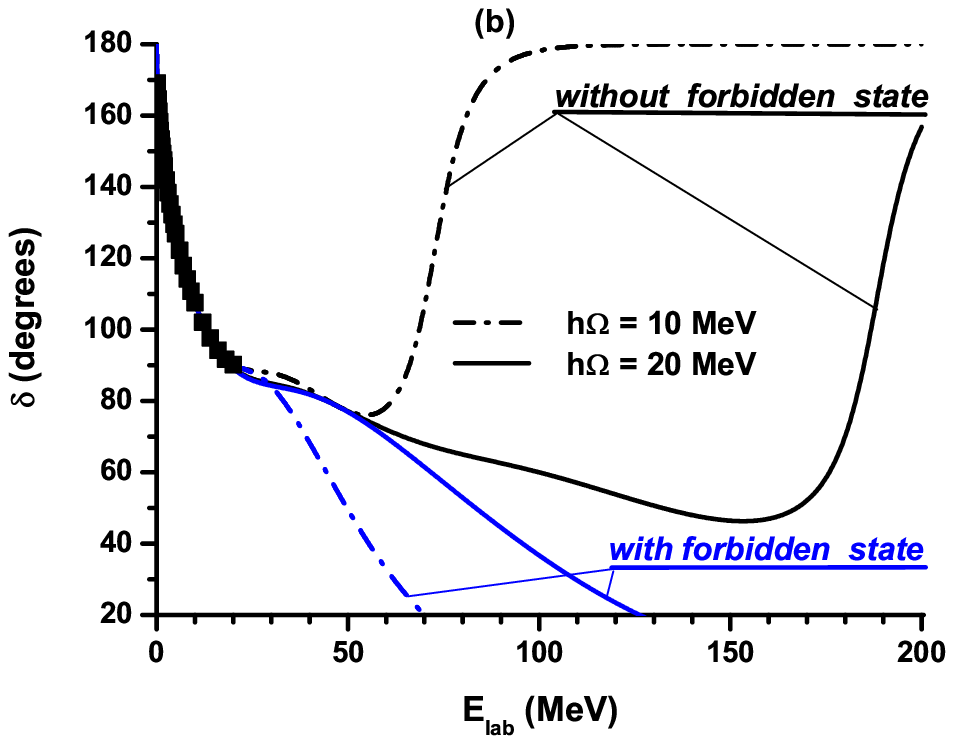,width=0.47\textwidth}}
\caption{(Color online) The $J$-matrix parameterization of the
 $\frac12^+$ $n\alpha$ phase shifts obtained in the $7\hbar\Omega$ model
 space with different values of oscillator spacing $\hbar\Omega$. See
 Fig.~\ref{phnap3n4} for details.}\vspace{-2ex}
\label{phnas1n4}
\end{figure}

In Fig. \ref{phnas1n4} we present the $J$-matrix parameterization of the
$\frac12^+$ phase shifts in elastic $n\alpha$ scattering in the
$7\hbar\Omega$ model space with different values of the oscillator
spacing $\hbar\Omega$. As usual, larger  $\hbar\Omega$ value makes it
possible to describe the phase shifts in a larger energy interval. A new
and interesting issue is the difference in behavior of the phase shifts in
the models with and without a forbidden state. A more realistic model
with forbidden state provides a proper dependence of the phase shifts:
starting with $180^\circ$ at zero energy, they tend to zero at large
energies. The forbidden state makes the same contribution to the
Levinson theorem as any other bound state providing the
$180^\circ$ difference between the phase shifts at zero
and infinite energies. The model without a forbidden state generates the
phase shifts returning at large energies back  to their zero energy
value. In what follows, we use the potential model with a forbidden
state. Note however that in the energy interval of known phase shifts, the
parameterizations of both models are indistinguishable. The $E_\lambda$
values provided by both models in this energy interval, are the same.

\begin{figure}
\centerline{\epsfig{file=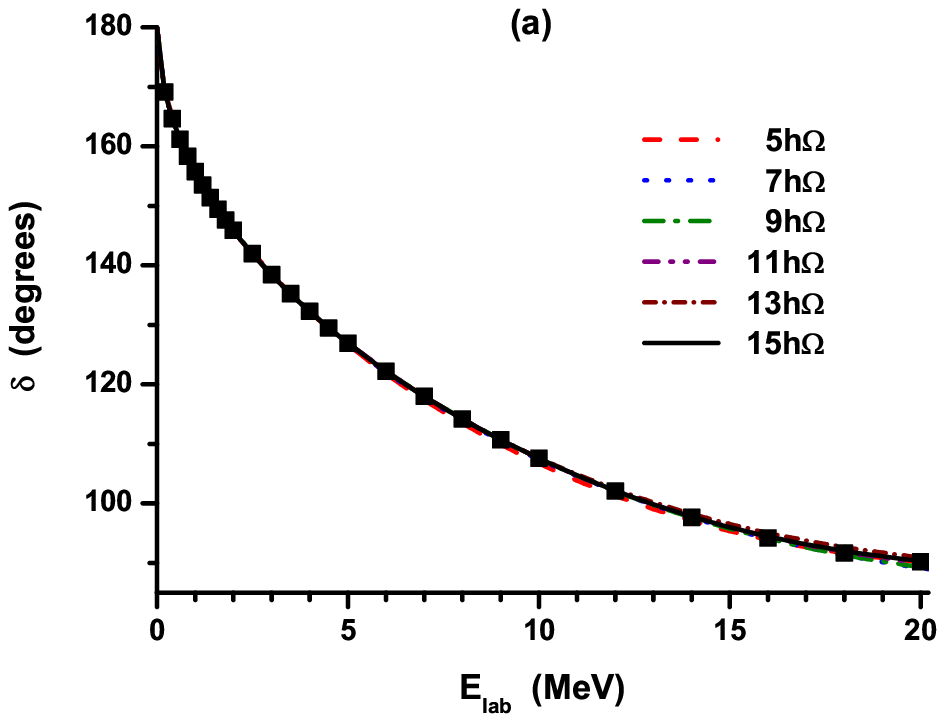,width=0.47\textwidth}}
\centerline{\epsfig{file=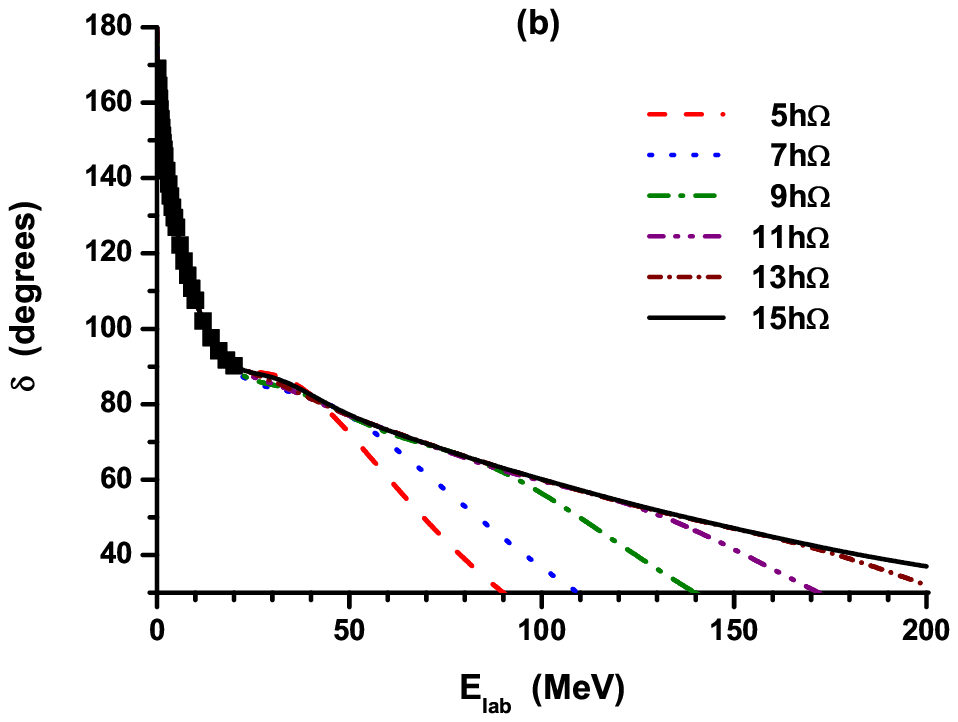,width=0.47\textwidth}}
\caption{(Color online) The $J$-matrix parameterization of the
 $\frac12^+$ $n\alpha$ phase shifts obtained  in the model with
 forbidden state  with $\hbar\Omega=20$ MeV
 in various model spaces.
 See Fig.~\ref{phnap3n4} for details.}
 \label{phnasfh20}
\end{figure}

The $\frac12^+$ phase shifts parameterizations in different model spaces
with $\hbar\Omega=20$ MeV perfectly describe the data
(Fig. \ref{phnasfh20}). At larger energies, they follow general trends:
smaller model spaces result in a faster fall off of the phase shifts to
zero value.

\section{\boldmath Analysis of $p\alpha$ scattering phase shifts}
\subsection{\boldmath $\frac32^-$ phase shifts}

The $J$-matrix approach to $p\alpha$ scattering
involves  an additional parameter
$b$, the channel radius used to define the auxiliary potential $V^{Sh}$
by truncating the Coulomb interaction at $r=b$ [see
Eq. (\ref{eq:pot1})]. We start our discussion of the $J$-matrix inverse
scattering description of $p\alpha$ scattering from the analysis of the
$b$-dependence of the $\frac32^-$ $p\alpha$ phase shift parameterization.

\begin{figure}
\centerline{\epsfig{file=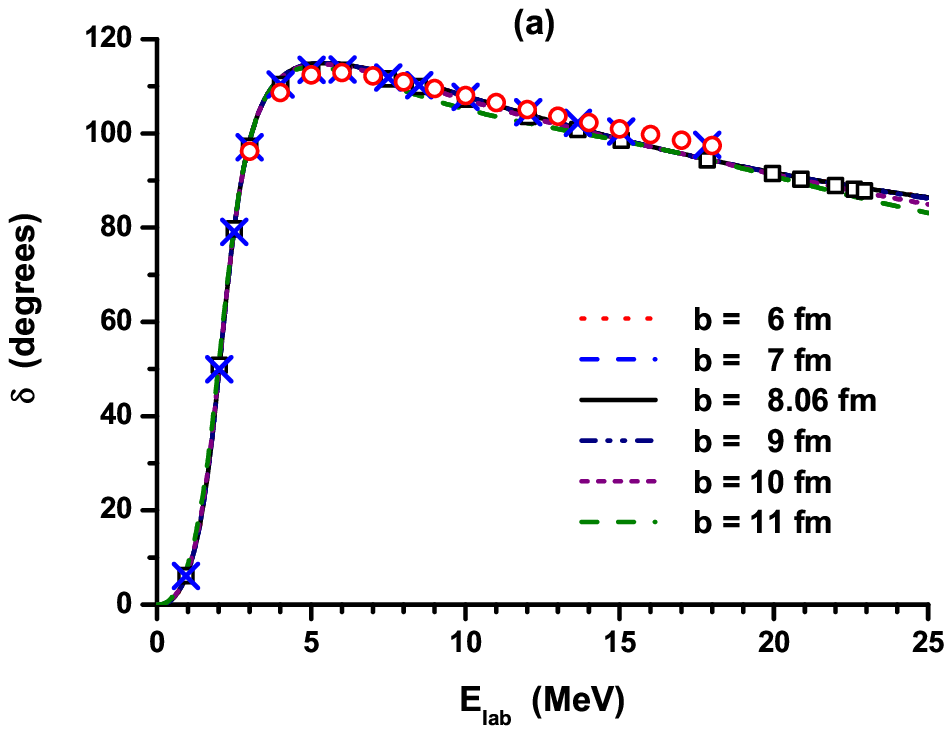,width=0.47\textwidth}}
\centerline{\epsfig{file=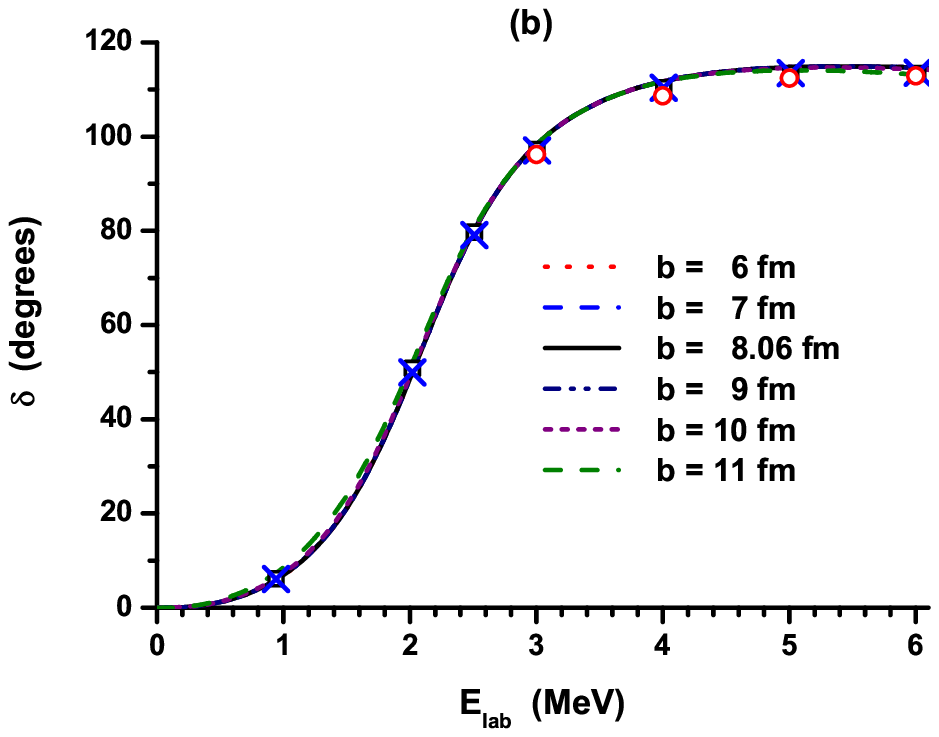,width=0.47\textwidth}}\vspace{-2ex}
\caption{(Color online) The $J$-matrix parameterization of the
 $\frac32^-$ $p\alpha$ phase shifts obtained with $\hbar\Omega=20$ MeV
 in the $10\hbar\Omega$ model space with various $b$ values. Two
 panels present the same results in
 different  scales. Experimental phase shifts:  open squares~---
 Ref. \cite{Arndt}, open circles~--- Ref. \cite{NPA163}, crosses~---
 Ref. \cite{Dodder}.}
\label{phpap3-b}
\end{figure}

\begin{figure}
\centerline{\epsfig{file=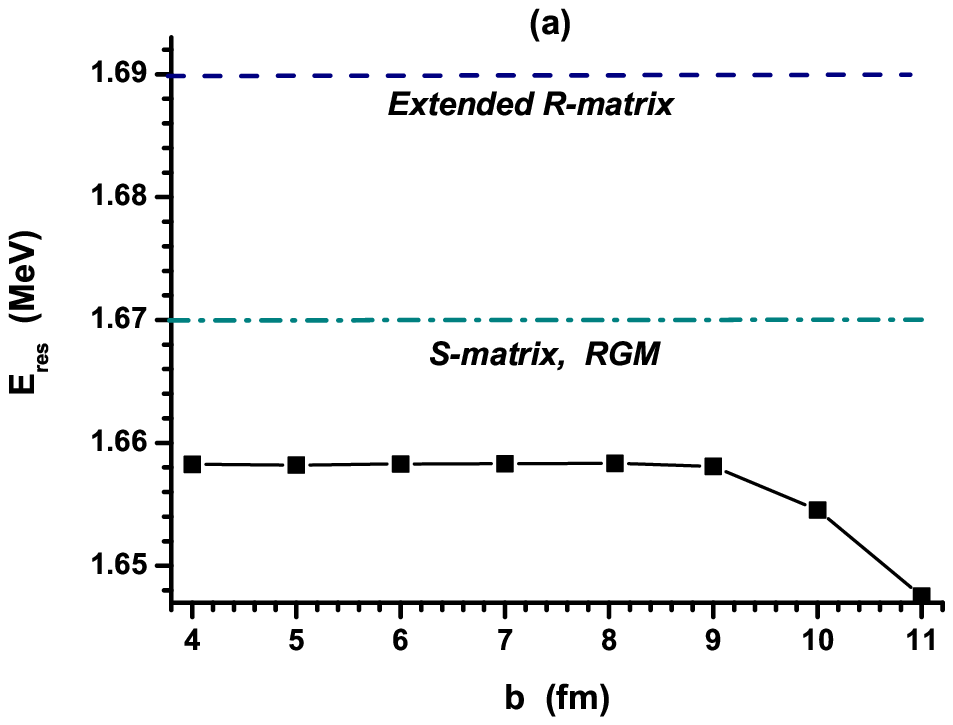,width=0.47\textwidth}}
\centerline{\epsfig{file=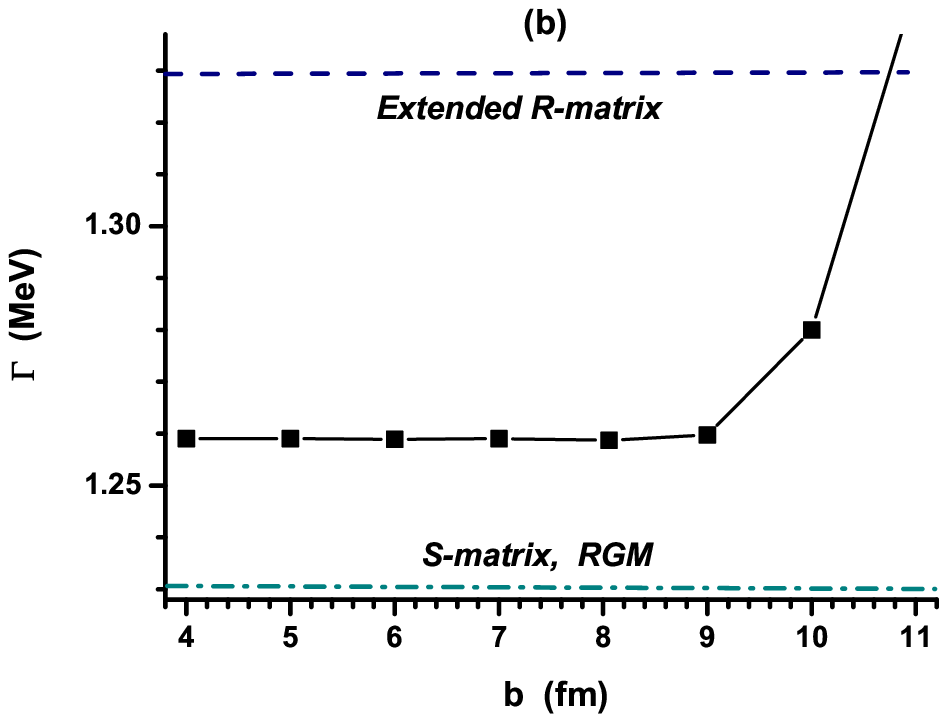,width=0.47\textwidth}}\vspace{-1.5ex}
\caption{(Color online) Dependence of the $\frac32^-$ $p\alpha$
 resonance energy  in
 the center-of-mass frame (upper panel) and width (lower panel) on the channel
 radius $b$ in the $10\hbar\Omega$ model space calculations with
 $\hbar\omega =20$ MeV. See Fig. \ref{ergamp3} for details.}\vspace{-1ex}
\label{pap3-erg-b}
\end{figure}

\begin{figure}
\centerline{\epsfig{file=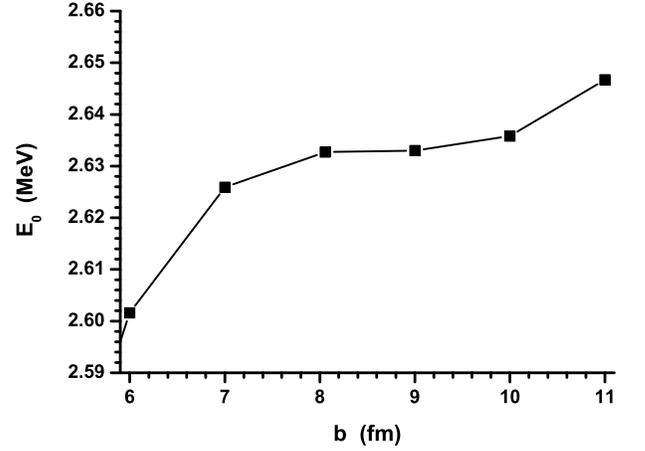,width=0.47\textwidth}}\vspace{-1.5ex}
\caption{
The $b$ dependence of the lowest state
 $E_{\lambda=0}$  in the $10\hbar\Omega$ model space $J$-matrix
 parameterizations with $\hbar\omega =20$ MeV. }\vspace{-1ex}
 \label{pap3-e0v2-b}
\end{figure}

We present in Fig. \ref{phpap3-b} the $p\alpha$ $\frac32^-$ phase shift
parameterizations obtained with different channel radii $b$ in the
$10\hbar\Omega$ model space  with $\hbar\Omega=20$ MeV. The
experimental data are seen to be perfectly described in the
interval of $b$ values $\rm 6~fm\leq b\leq 10~fm$. However we did not
find a way to reproduce accurately the phase shifts with $b\geq 11$~fm,
in particular at energies between 10 and 20 MeV. This is not surprising
since the classical turning point of the highest oscillator function
$R_{{\cal N}l}(r)$ involved in the construction of the truncated
Hamiltonian $H^l_{nn'}$ $(n,n'\leq{\cal N})$, $r^{cl}_{\cal N}=8.06$~fm
in this case. The $\frac32^-$
resonance energy
and width dependences on $b$ obtained from these parameterizations, are
shown in Fig. \ref{pap3-erg-b}. The resonance parameters are seen to be
stable enough with $b$ varying between 6 and 10 fm. The $b$ dependence
of the energy $E_{\lambda=0}$ of the lowest state obtained in the
$J$-matrix parameterization, has a plateau between 7 and 10 fm (see
Fig. \ref{pap3-e0v2-b}).

The bottom line of these studies is that the results are nearly
$b$-independent for $b$ values in some vicinity of the classical turning
point $r^{cl}_{\cal N}$. This conclusion remains valid for other
partial waves of the $p\alpha$ scattering and we are not discussing
$b$-dependences in the following subsections.
The
remainder of the calculations presented here are performed with $b=r^{cl}_{\cal N}$.

\begin{figure}
\centerline{\epsfig{file=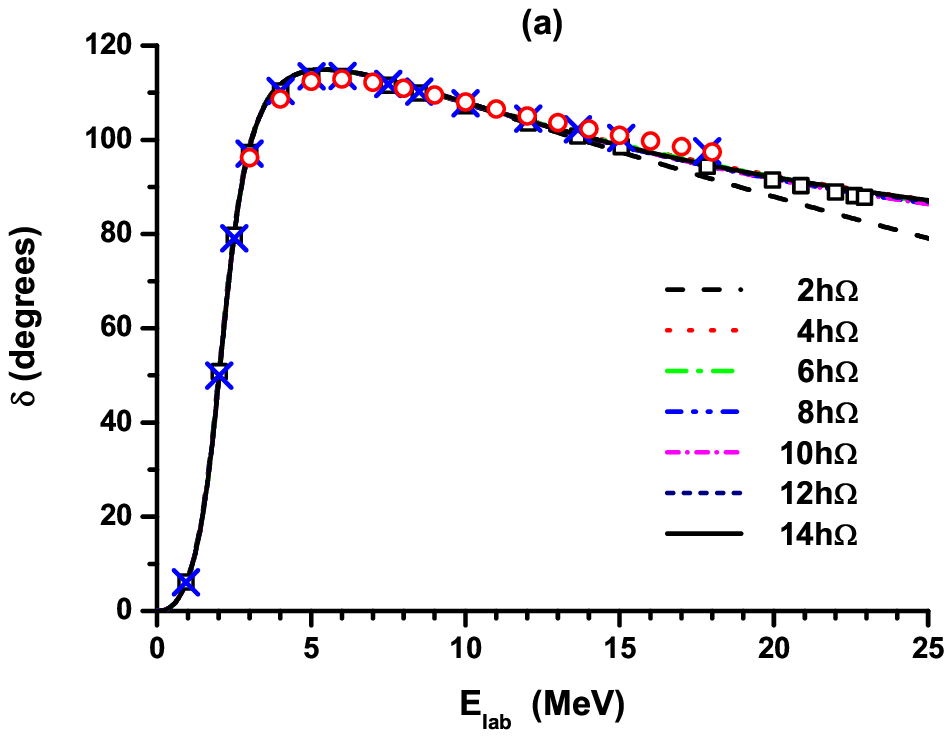,width=0.47\textwidth}}
\centerline{\epsfig{file=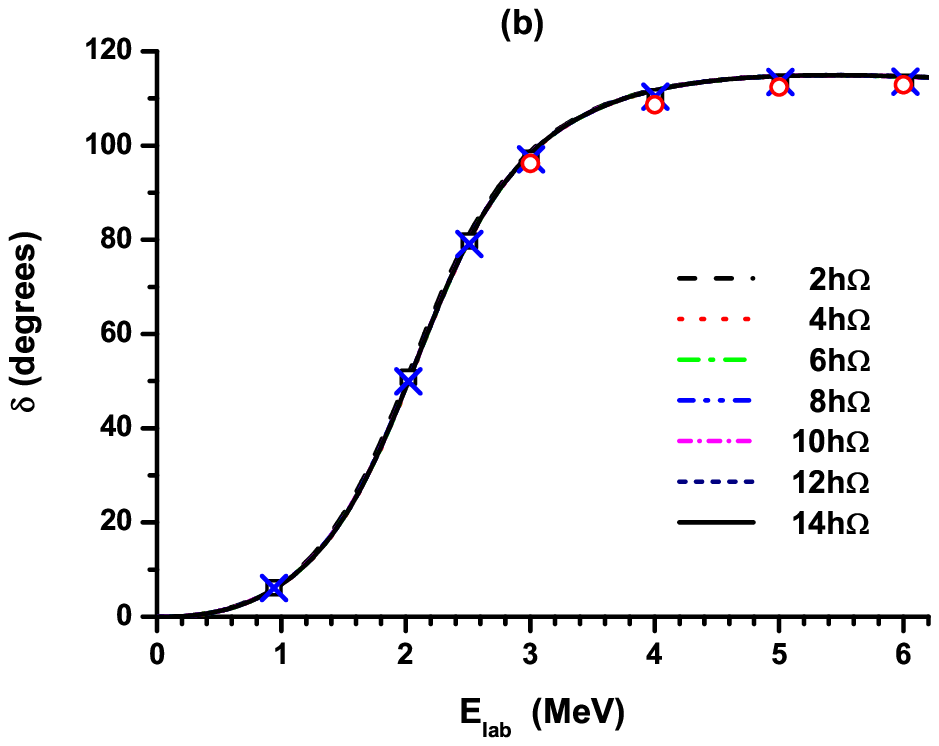,width=0.47\textwidth}}\vspace{-1.5ex}
\caption{(Color online) The $J$-matrix parameterization of the
$p\alpha$
 $\frac32^-$  phase shifts obtained with $\hbar\Omega=20$ MeV
 in various model spaces. See Fig. \ref{phpap3-b} for details.}
\label{phpap3h20}
\end{figure}

\begin{figure*}
{\epsfig{file=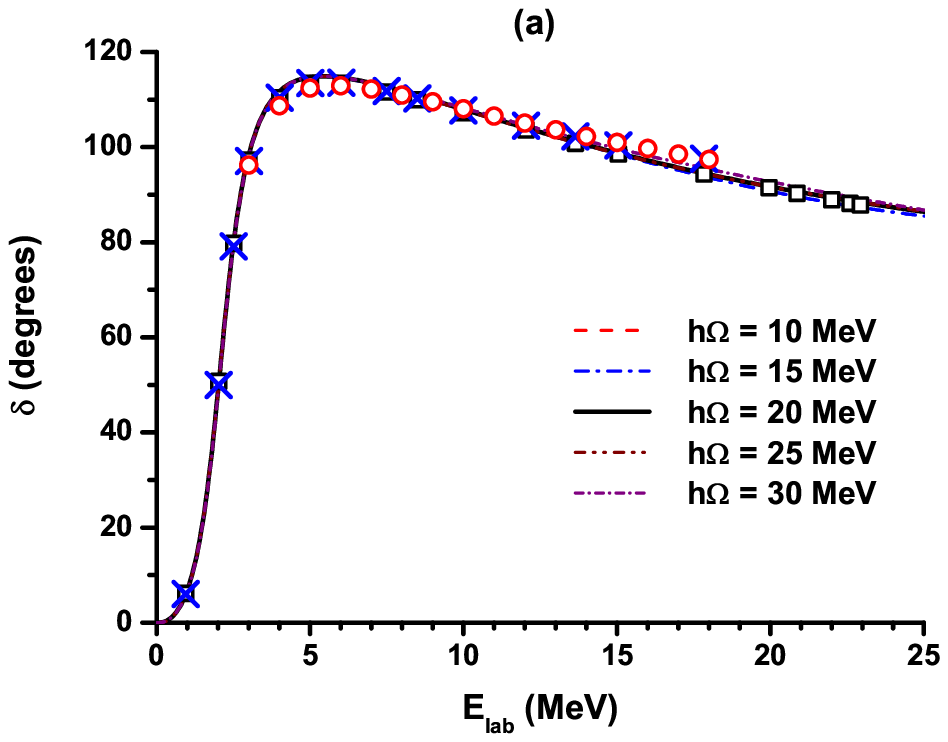,width=0.47\textwidth}}\hfill
{\epsfig{file=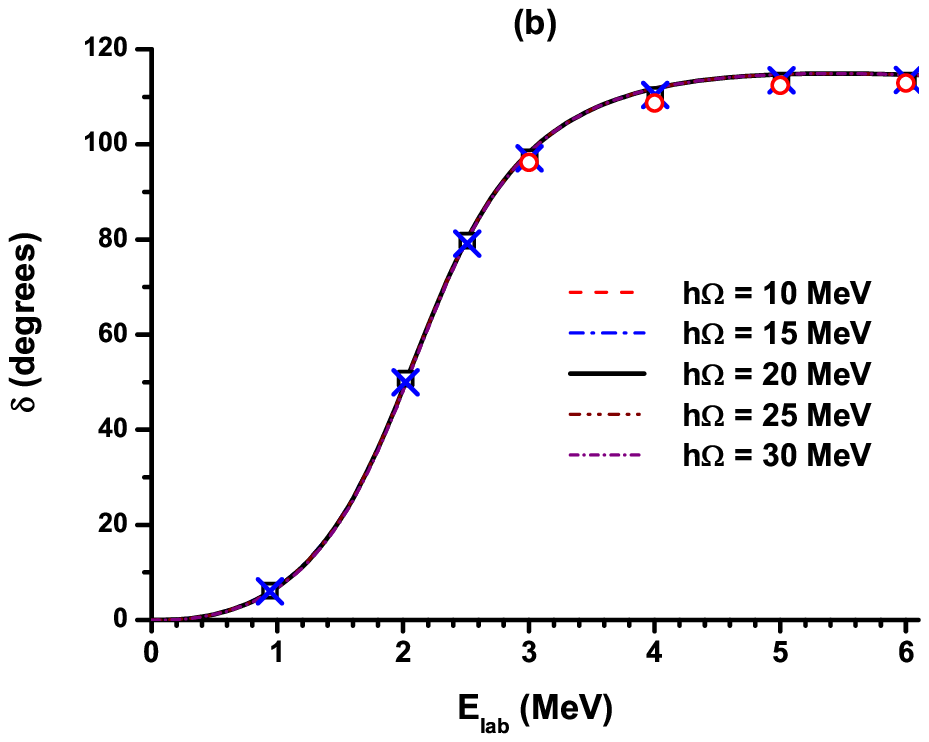,width=0.47\textwidth}}\vspace{-1.5ex}
\caption{(Color online) The $J$-matrix parameterization of the
$p\alpha$
 $\frac32^-$  phase shifts obtained in the $10\hbar\Omega$ model space
 with various $\hbar\Omega$ values. See Fig. \ref{phpap3-b} for details.}
\label{phpap3N10}
\end{figure*}

\begin{figure*}
{\epsfig{file=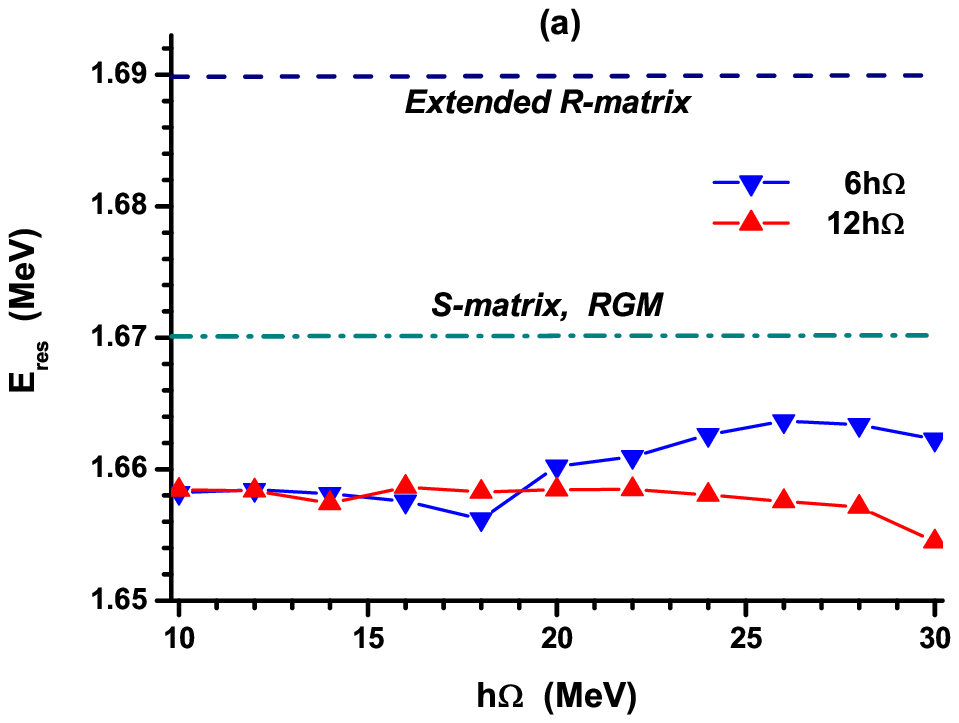,width=0.47\textwidth}}\hfill
{\epsfig{file=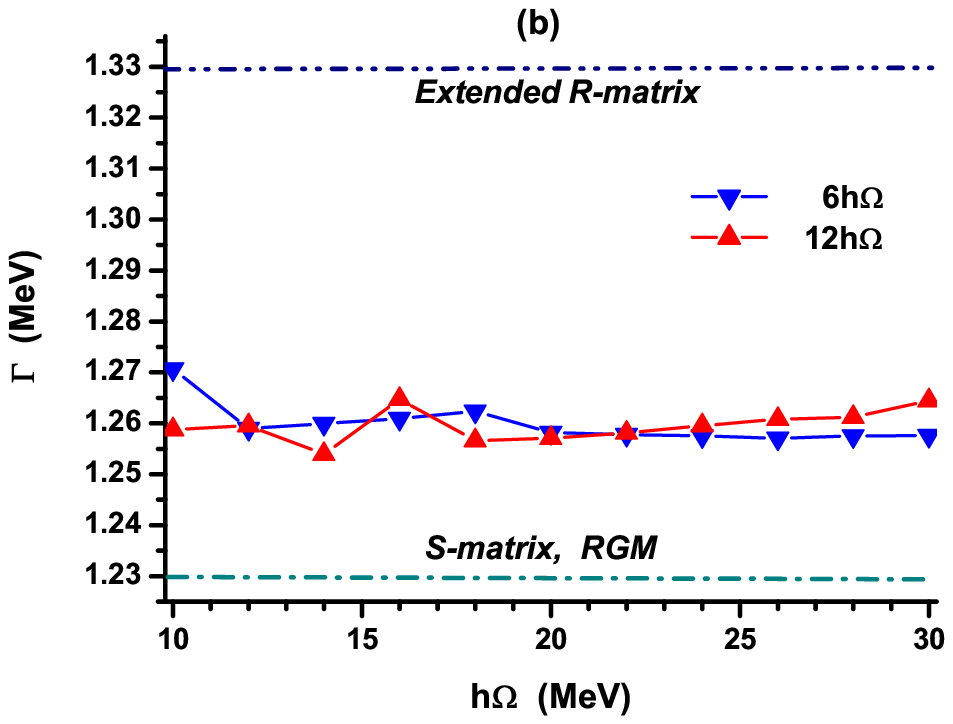,width=0.47\textwidth}}
{\epsfig{file=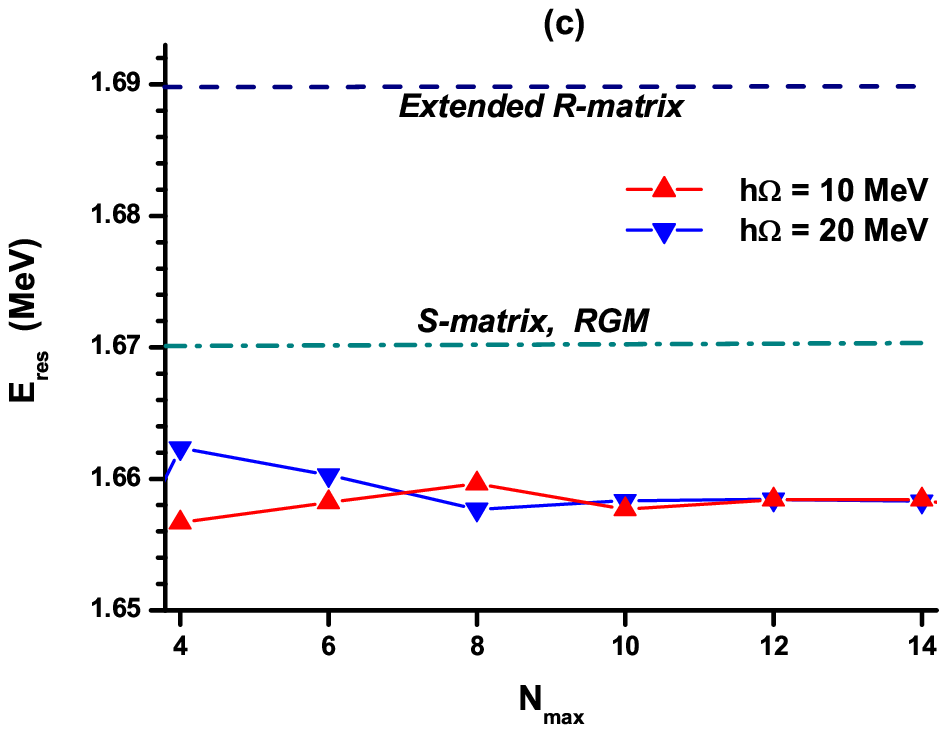,width=0.47\textwidth}}\hfill
{\epsfig{file=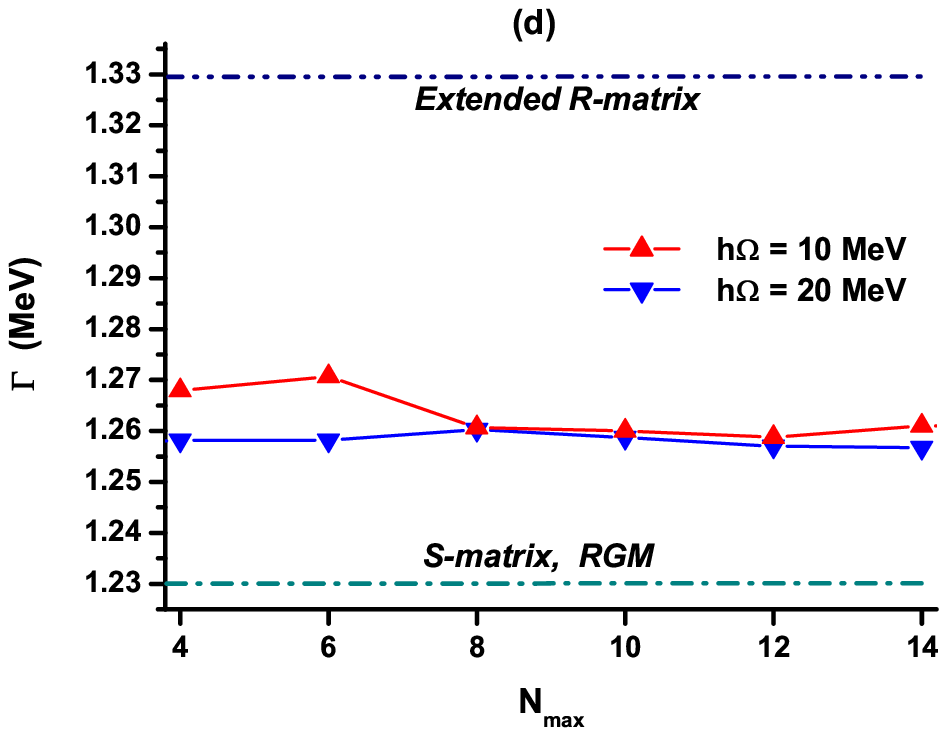,width=0.47\textwidth}}
\caption{(Color online) The  $p\alpha$ $\frac32^-$ resonance energy
in
 the center-of-mass frame
 (left) and width (right) obtained by calculating the position of the
 $S$-matrix pole by means of the $J$-matrix parameterizations with
 different $\hbar\Omega$ values (upper panels) and in different model spaces
(lower panels). See Fig. \ref{ergamp3} for details.}
\label{pap-ergamp3}
\end{figure*}

We present in Fig. \ref{phpap3h20} the $J$-matrix parameterization of
the $\frac32^-$ $p\alpha$ phase shifts obtained  in various model spaces
with $\hbar\Omega=20$ MeV. The data are well-described in $4\hbar\Omega$
and higher model spaces. Some deviation from experiment is seen only for
the $2\hbar\Omega$ model space starting from laboratory energies about
20 MeV. However, the resonance region is perfectly described even in
this very small $2\hbar\Omega$ model space as is seen from the lower
panel of Fig. \ref{phpap3h20} where the enlarged energy scale is
used. The $J$-matrix parameterization  is also insensitive to the
variation of the $\hbar\Omega$ value in the whole interval of known
phase shifts  including the resonance region (see
Fig. \ref{phpap3N10}). Therefore it is not surprising that we obtain
a very stable description of the resonance energy and width
(see Fig. \ref{pap-ergamp3}), one that is independent
of the model space and $\hbar\Omega$ value.

Our results for the $\frac32^-$ resonance parameters are very close to
the ones obtained in the analysis of Ref. \cite{Rmatr}.

\begin{figure}
\centerline{\psfig{figure=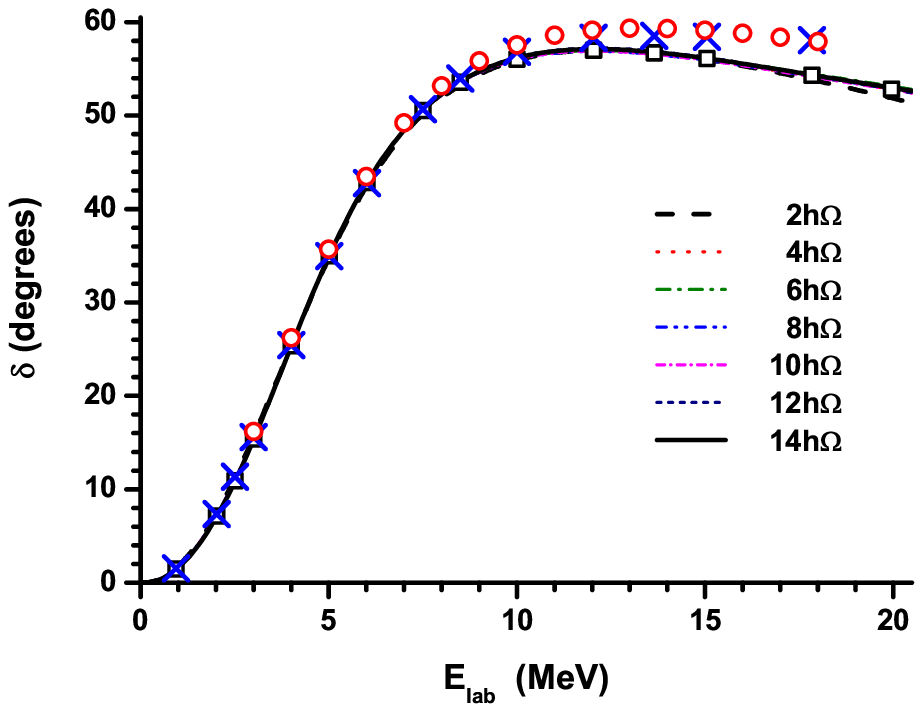,width=0.47\textwidth}}
\caption{(Color online) The $J$-matrix parameterization of the $p\alpha$
 $\frac12^-$  phase shifts obtained with $\hbar\Omega=20$ MeV
 in various model spaces. See Fig. \ref{phpap3-b} for details.}
\label{phpap1h20}
\end{figure}

\begin{figure}
\centerline{\psfig{figure=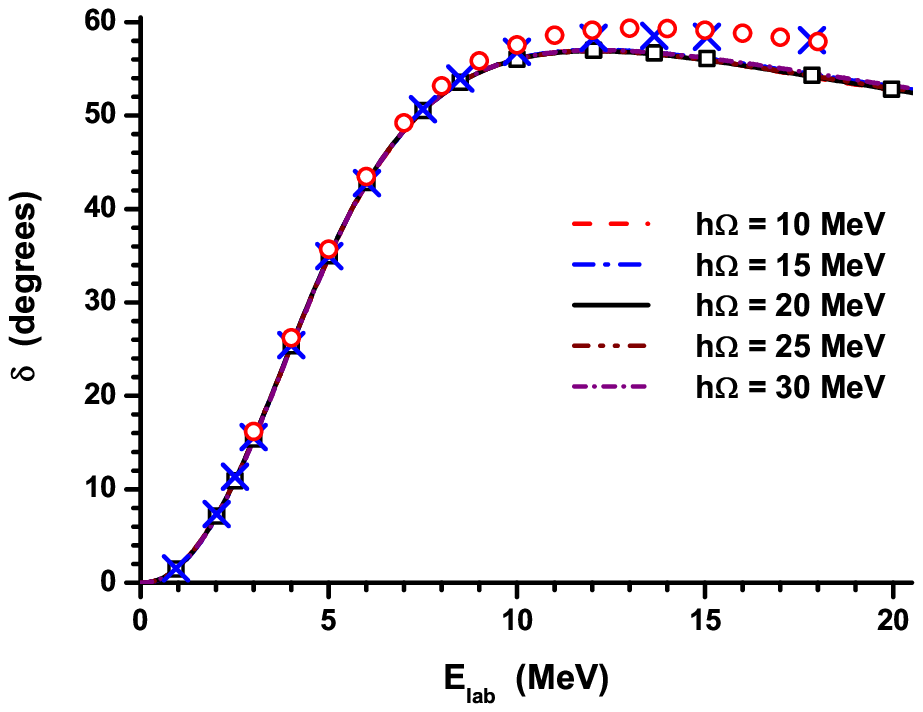,width=0.47\textwidth}}
\caption{(Color online) The $J$-matrix parameterization of the $p\alpha$
 $\frac12^-$  phase shifts obtained in the $10\hbar\Omega$ model space
 with various $\hbar\Omega$ values. See Fig. \ref{phpap3-b} for details.}
\label{phpap1N10}
\end{figure}

\begin{figure*}
{\epsfig{file=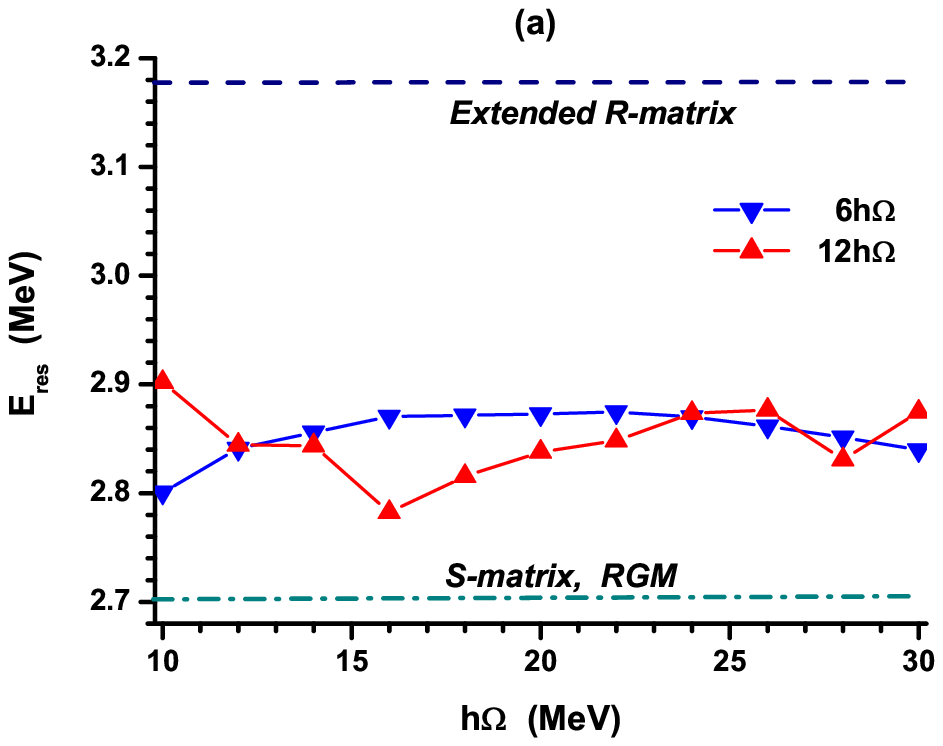,width=0.47\textwidth}}\hfill
{\epsfig{file=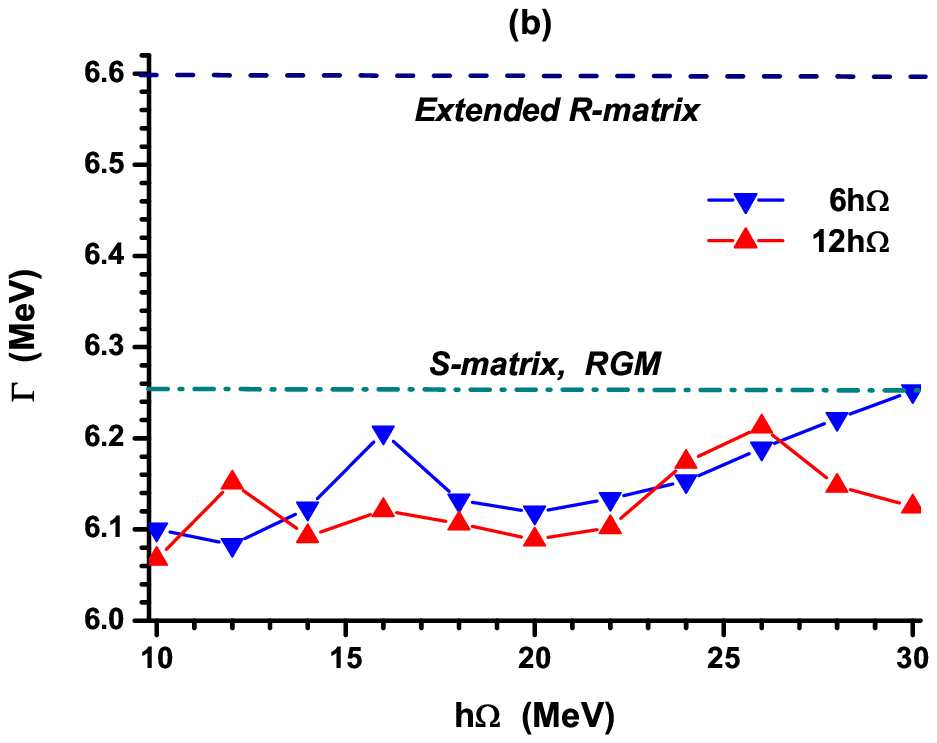,width=0.47\textwidth}}
{\epsfig{file=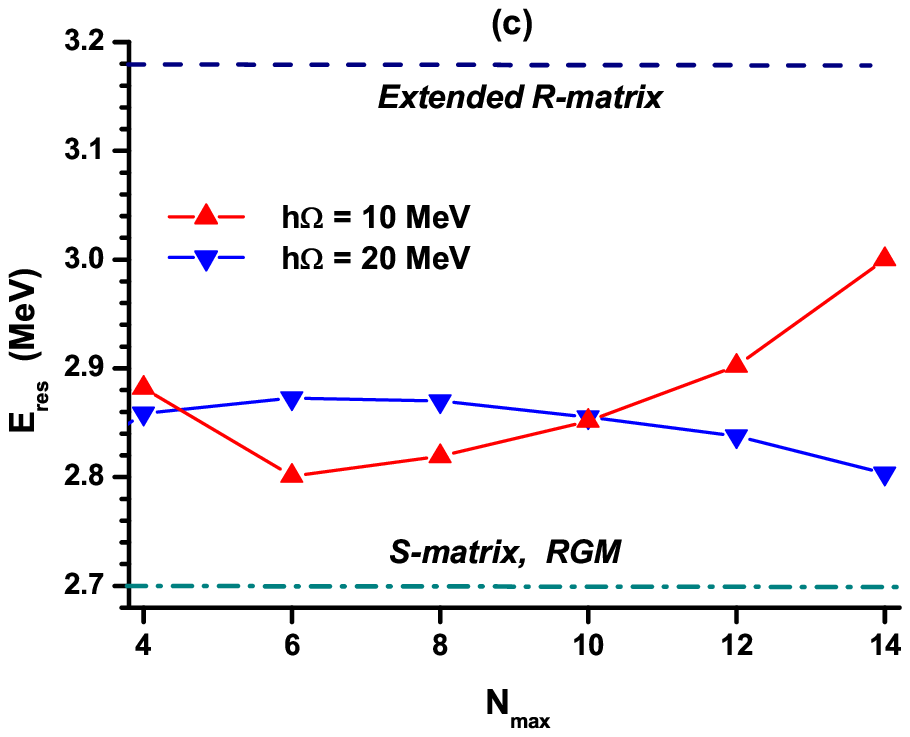,width=0.47\textwidth}}\hfill
{\epsfig{file=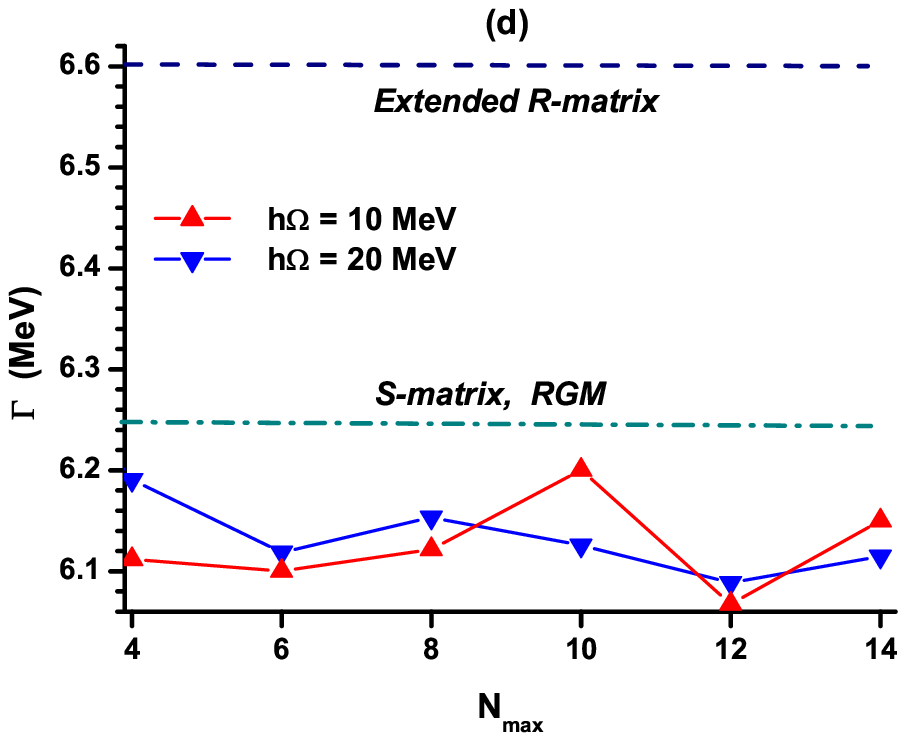,width=0.47\textwidth}}\vspace{-1.5ex}
\caption{(Color online) The  $p\alpha$ $\frac12^-$ resonance energy in
 the center-of-mass frame
 (left) and width (right) obtained by calculating the position of the
 $S$-matrix pole by means of the $J$-matrix parameterizations with
 different $\hbar\Omega$ values (upper panels) and in different model spaces
(lower panels). See Fig. \ref{ergamp3} for details.}
\label{pap-ergamp1}
\end{figure*}

\subsection{\boldmath $\frac12^-$ phase shifts}

We obtain a high-quality $J$-matrix parameterization of the $p\alpha$
$\frac12^-$ phase shifts, very stable with variations of the model space
or oscillator spacing $\hbar\Omega$.  A small deviation from the
experiment at large energies is seen
in Fig. \ref{phpap1h20} in the $2\hbar\Omega$ model space only. The
parameterizations obtained in the $10\hbar\Omega$ model space with
$\hbar\Omega$ values ranging from 10 to 30 MeV, are indistinguishable in
Fig. \ref{phpap1N10}. The resonance region is perfectly described.
Our results for the resonance energy and width correspond well to the
analysis of Ref. \cite{Rmatr}. The resonance parameters are stable with
respect to variations of the model space and $\hbar\Omega$ (see
Fig. \ref{pap-ergamp1}). Of course, the variations of $E_{res}$ and
$\Gamma$ in Fig. \ref{pap-ergamp1} are much larger than in the case of
the $\frac32^-$ resonance, but the $\frac12^-$ resonance energy and
width are also much larger
than the energy and width of the $\frac32^-$ resonance.

\subsection{\boldmath $\frac12^+$ phase shifts}

\begin{figure*}
{\psfig{figure=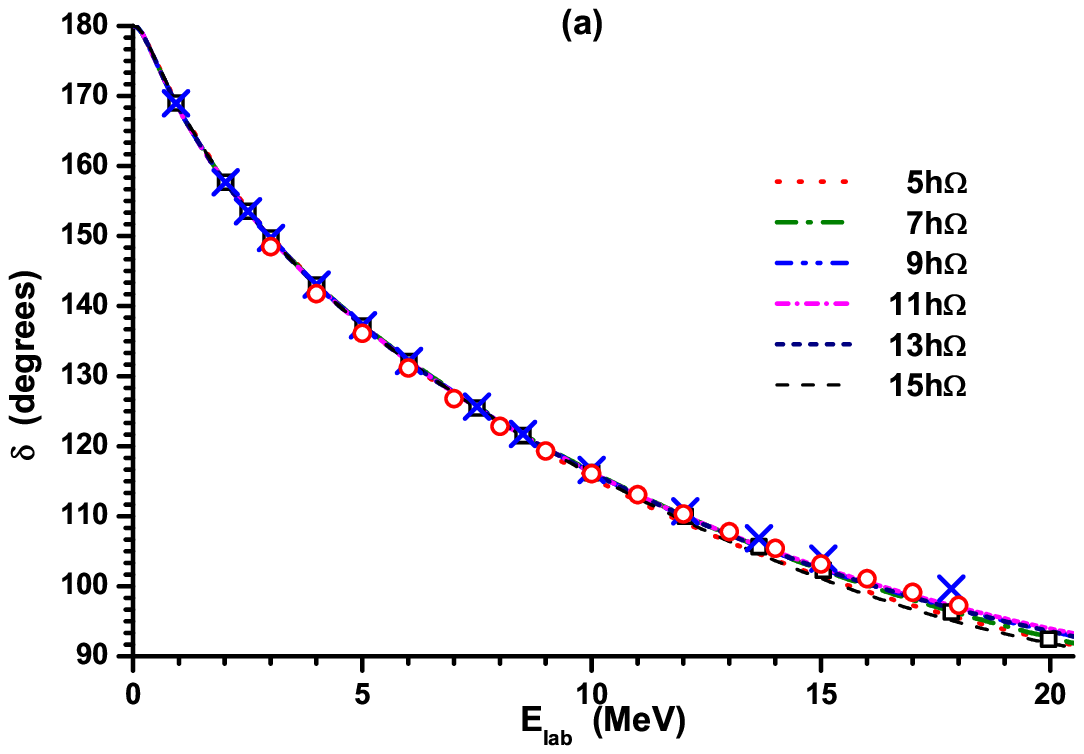,width=0.47\textwidth}}\hfill
{\psfig{figure=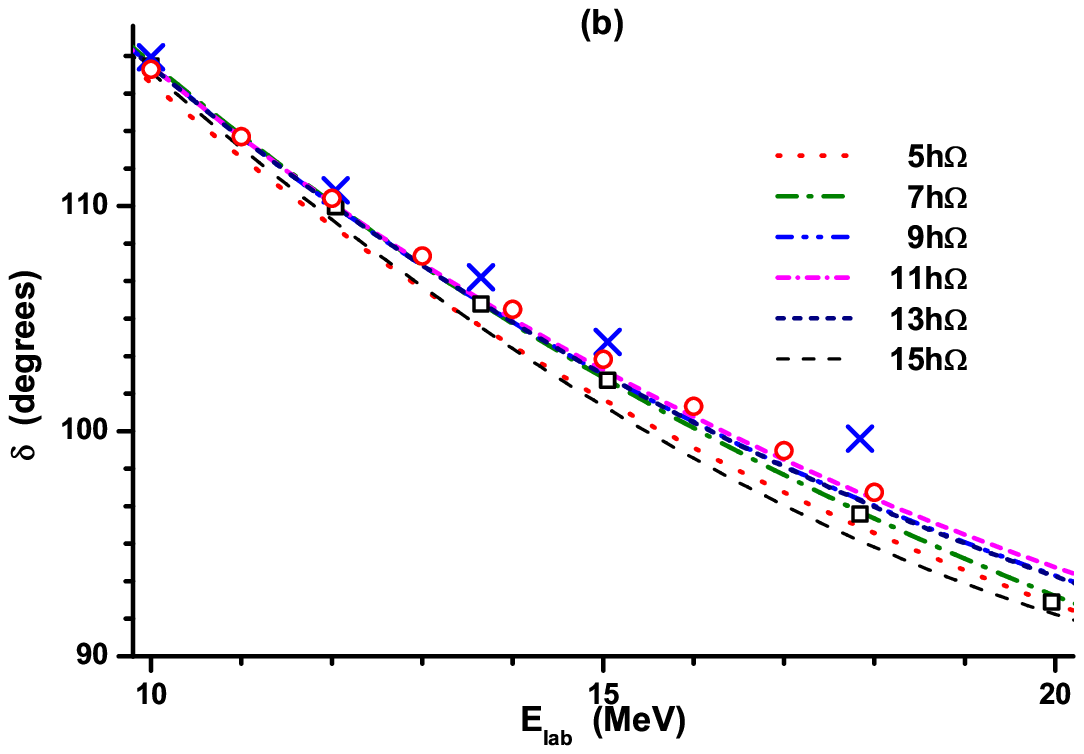,width=0.47\textwidth}}
\caption{(Color online) The $J$-matrix parameterization of the $p\alpha$
 $\frac12^+$  phase shifts obtained in the model with forbidden state
with $\hbar\Omega=20$ MeV in various model spaces. See
 Fig. \ref{phpap3-b} for details.}
\label{phpasfh20}
\end{figure*}

In the case of $s$ wave of $p\alpha$ scattering, we can also use
interaction models
with and without a forbidden state. The main features of the $J$-matrix
parameterizations within these models in the case of the $p\alpha$
scattering are the same as in the case of $n\alpha$ scattering; in
particular, the phase shift description in the  low-energy region
covering the whole region of known phase shifts, is identical within
these interaction models. In what follows, we present only the results
obtained in the model with forbidden state which we suppose to be more
realistic.

The $J$-matrix parameterizations of the $p\alpha$ $\frac12^+$ phase
shifts obtained in various model spaces with $\hbar\Omega=20$ MeV, are
presented in Fig. \ref{phpasfh20} in two scales. The low-energy phase
shifts up to approximately $E_{lab}=10$ MeV are perfectly reproduced in
all model spaces. Starting from $E_{lab}=10$ MeV, there are some
deviations from the experiment that are well seen in the 
right panel of Fig. \ref{phpasfh20} where
a larger scale is used. Surprisingly, the
deviations from experimental phase shifts are larger in larger model
spaces. The deviations are not  large but not negligible.

\begin{figure}
\centerline{\psfig{figure=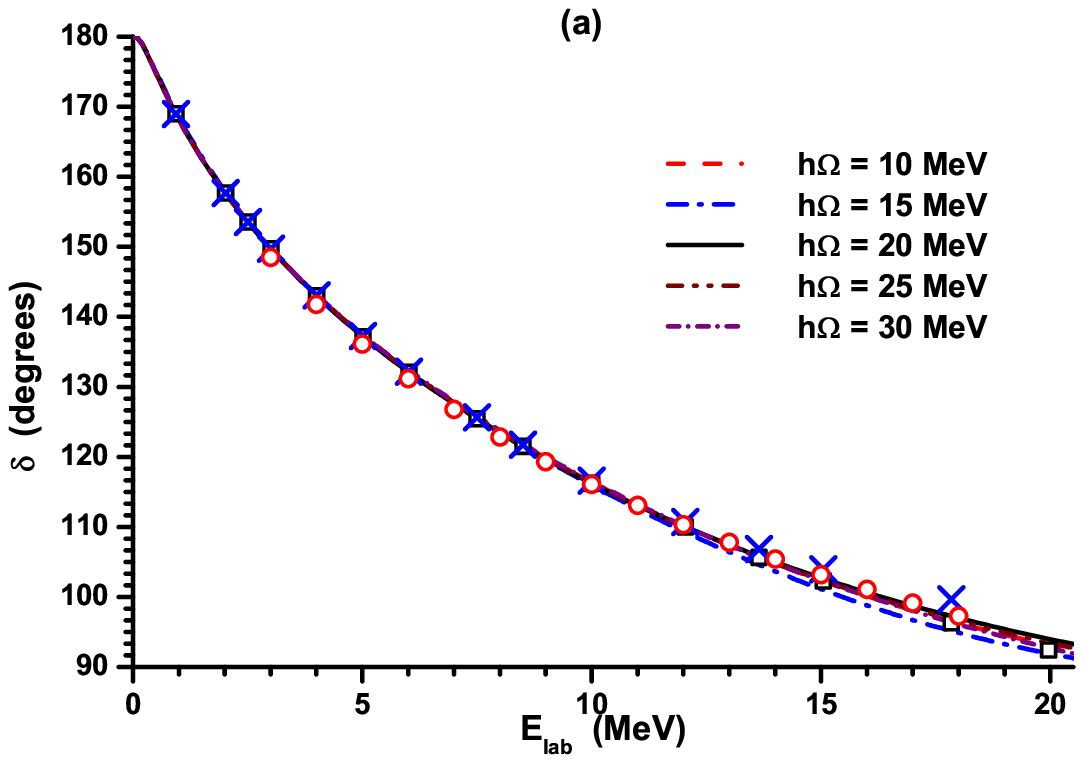,width=0.47\textwidth}}
\centerline{\psfig{figure=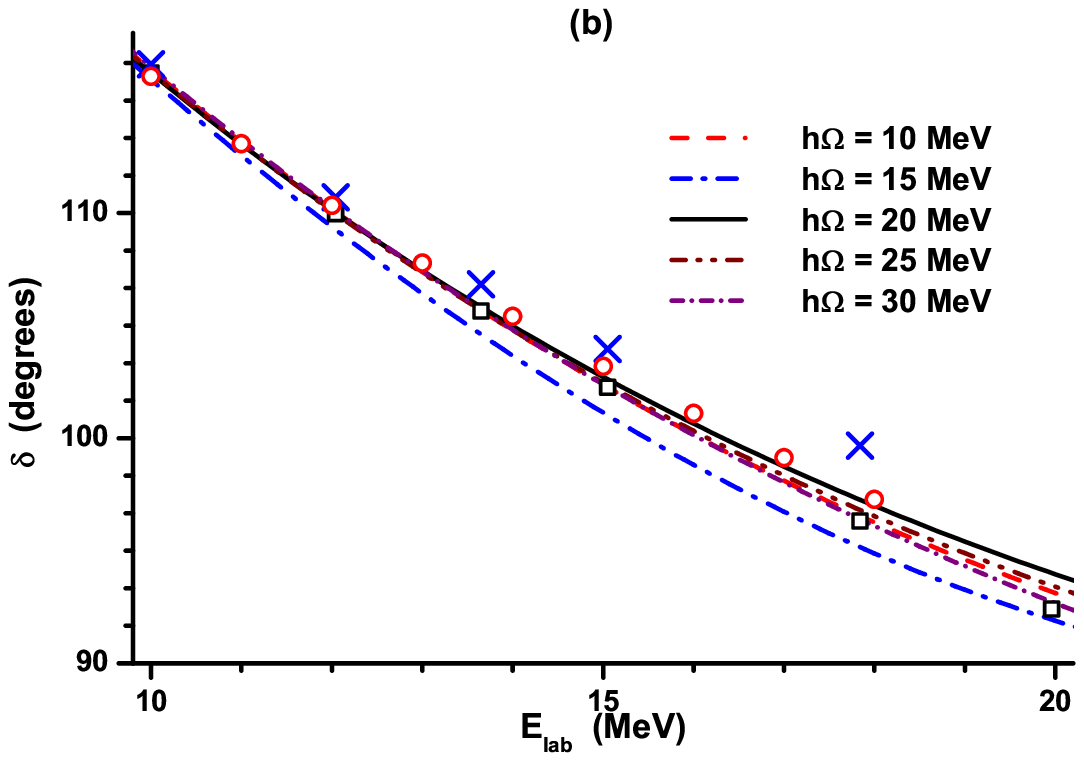,width=0.47\textwidth}}
\caption{(Color online) The $J$-matrix parameterization of the $p\alpha$
 $\frac12^+$  phase shifts obtained  in the model with forbidden state
in the $11\hbar\Omega$ model space
 with various $\hbar\Omega$ values. See Fig. \ref{phpap3-b} for details.}
 \label{phpasfN11}
\end{figure}

The  $J$-matrix parameterizations obtained with various $\hbar\Omega$
values in the $11\hbar\Omega$ model space, are shown in
Fig. \ref{phpasfN11}. The theoretical curves are nearly
indistinguishable below $E_{lab}=10$ MeV reproducing well
the experimental
data. Some difference between parameterizations is seen in the
high-energy part of the interval of known phase shifts.
All $J$-matrix
parameterizations presented in Fig. \ref{phpasfN11} reasonably describe
the phenomenological data in the whole energy interval  of known phase
shifts. The worst description of the phase shifts in the $11\hbar\Omega$
model space is obtained with $\hbar\Omega=15$  MeV.

\section{\boldmath $J$-matrix and shell model eigenstates}\label{SM-El}

Up to now, we were discussing the $J$-matrix inverse scattering
description of scattering observables in the $n+\alpha$ and $p+\alpha$
nuclear systems. It is very interesting to investigate whether these observables
correlate with  the shell model predictions for $^5$He and
$^5$Li nuclei. It should be
done, as we have shown above, by comparing the eigenenergies $E_\lambda$
obtained in the $J$-matrix inverse scattering approach with the energies
of the states obtained in the shell model.

We calculate the lowest $^5$He and $^5$Li states of a given spin and
parity  in the No-core Shell Model  approach
\cite{Vary} using the code MFDn \cite{Vary92_MFDn} and the JISP16
nucleon-nucleon interaction
\cite{JISP16,JWEB}.
We do not make use of effective interactions
calculated within Lee--Suzuki or any other approach.  That is, all results
presented here are obtained with
the `bare' JISP16 $NN$ interaction which
is known \cite{JISP16,JISPYaF,extrap08} to provide a reasonable convergence
as basis space size increases.
One may note that the No-Core Shell Model with a bare interaction and with
a truncated configuration basis may also be referred to as a
``configuration interaction'' or ``CI'' type calculation \cite{Roth}.

In all cases, the calculations of the $^4$He ground
state energy is performed with the same $\hbar\Omega$ value and in the
same $N_{\max}\hbar\Omega$ model space. These $^4$He ground state
energies are used to calculate the reaction threshold while comparing
the $J$-matrix $E_\lambda$ values (defined with regard to the reaction
threshold) with the shell model results. Therefore our reaction
threshold is model space and $\hbar\Omega$-dependent, however these
dependencies are strongly suppressed in large enough model spaces. This
definition of the reaction threshold is, of course, somewhat
arbitrary. We use it supposing that our definition provides a
consistent way to generate energies relative to the $^4$He ground state
energy within the  No-core Shell Model  approach
employing a finite basis.

\begin{figure}
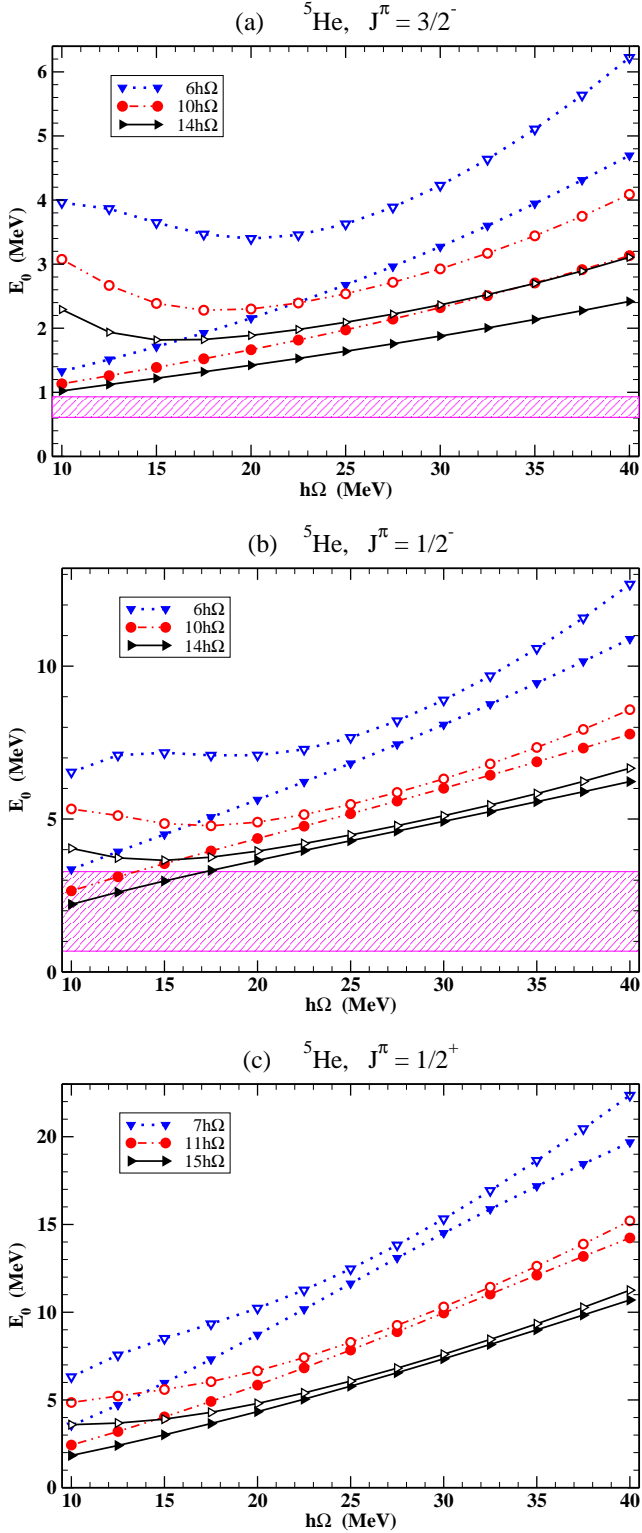

\centerline{\epsfig{file=nap3.eps,width=0.47\textwidth}}\vspace{2ex}
\centerline{\epsfig{file=nap1.eps,width=0.47\textwidth}}\vspace{2ex}
\centerline{\epsfig{file=nas1.eps,width=0.47\textwidth}}
\caption{(Color online) $E_{\lambda=0}$ values for $n\alpha$ scattering
 obtained in the $J$-matrix inverse scattering approach  in
 the center-of-mass frame (filled symbols) and respective lowest eigenstates of
the $^5$He nucleus  obtained in the No-core Shell Model
(respective empty
 symbols). The resonance energies together with their  widths are shown
 by shaded areas.}
\label{5He}
\end{figure}

\begin{figure}
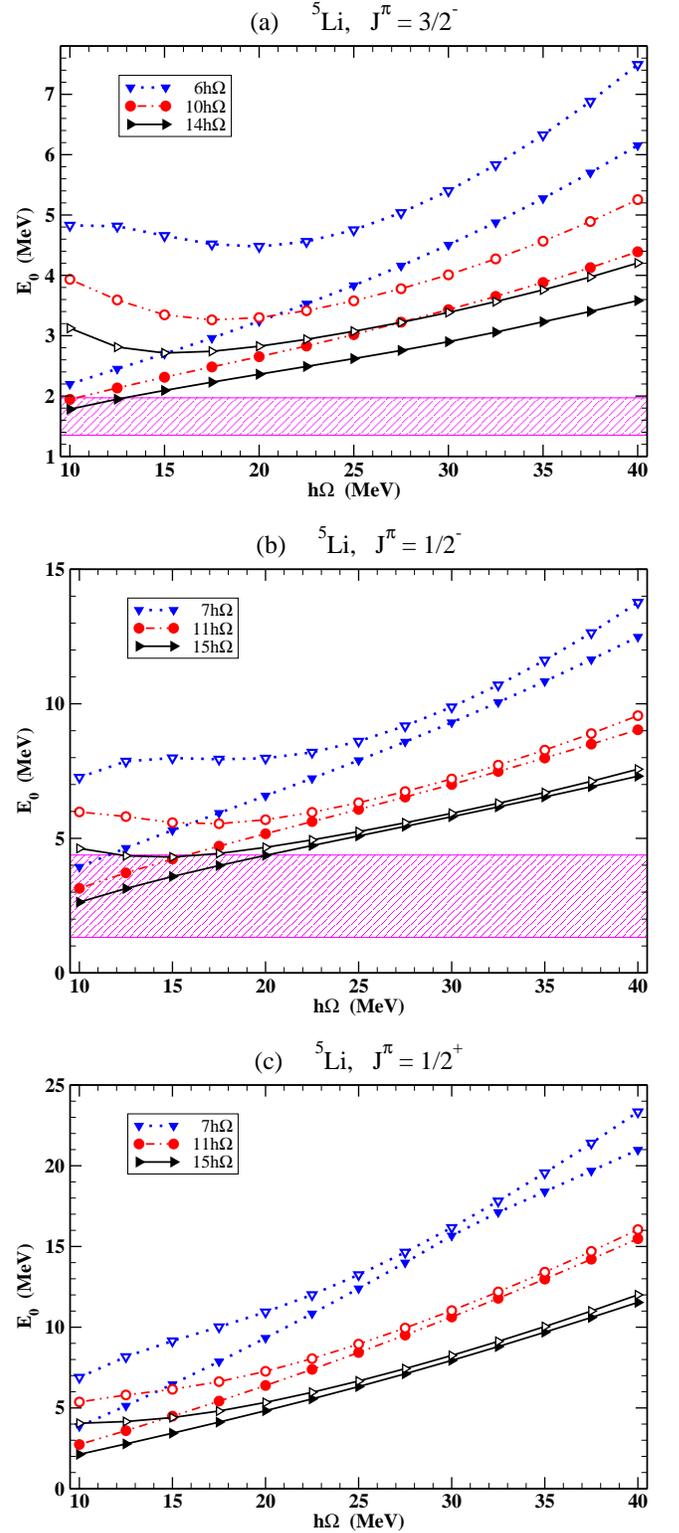

\centerline{\epsfig{file=pap3.eps,width=0.47\textwidth}}\vspace{2ex}
\centerline{\epsfig{file=pap1.eps,width=0.47\textwidth}}\vspace{2ex}
\centerline{\epsfig{file=pas1.eps,width=0.47\textwidth}}
\caption{(Color online) $E_{\lambda=0}$ values for $p\alpha$ scattering
 obtained in the $J$-matrix inverse scattering approach  in
 the center-of-mass frame
and respective lowest eigenstates of the $^5$Li nucleus  obtained in the
No-core Shell Model.
See Fig. \ref{5He} for details.}
\label{5Li}
\end{figure}

The No-core Shell Model
results for the lowest $^5$He and $^5$Li $\frac32^-$,
$\frac12^-$ and $\frac12^+$ states are compared with the respective $J$-matrix
$E_{\lambda=0}$ values  in Figs. \ref{5He} and
\ref{5Li}. For each spin and parity, the $J$-matrix $E_{\lambda=0}$ values obtained
with the same $\hbar\Omega$ value, are seen to decrease with increasing
model space (see also Fig. \ref{Elam0}); the same model space dependence
is well-known to be inherent for the shell model eigenstates. However
the $\hbar\Omega$ dependences of the $J$-matrix $E_{\lambda=0}$ and
shell model eigenstates, are very different: the shell model eigenstates
are known to have a minimum at some  $\hbar\Omega$ value while the
inverse scattering  $E_{\lambda=0}$ are seen from Figs. \ref{5He} and
\ref{5Li} to increase nearly linearly with $\hbar\Omega$; the slope of
the $\hbar\Omega$ dependence of $E_{\lambda=0}$ is larger for wider
resonances. As a result,
the shell model predictions differ from the results of the inverse
scattering analysis for small enough $\hbar\Omega$ values. However, a
remarkable correspondence between the shell model and inverse scattering
results is seen at large enough $\hbar\Omega$ values  starting from
approximately  $\hbar\Omega=20$ MeV.  The agreement between the shell
model and $J$-matrix inverse scattering analysis is improved with
increasing model space; it is probable that this is partly due to the
improvement in larger model spaces of the calculated threshold energy in
our approach. The shell model description of the lowest $\frac12^-$ and
$\frac12^+$ states is somewhat better than the lowest $\frac32^-$ state
description
in both $^5$He and $^5$Li nuclear systems. The lowest $\frac32^-$ state
description is however not so bad (note a more detailed energy scale for
the  $\frac32^-$ state in Figs. \ref{5He} and
\ref{5Li}): the difference between the shell model predictions and the
$J$-matrix analysis results is about 0.5 MeV in large enough model
spaces and for large enough $\hbar\Omega$ values. An excellent
description of the  $\frac12^-$ states in $^5$He and $^5$Li combined
with some deficiency in description of the  $\frac32^-$ states in the
same nuclei, is probably a signal of a somewhat underestimated strength
of the spin-orbit interaction generated by the JISP16 $NN$ interaction
in the $p$ shell.

We suppose that the results presented here illustrate well the power of
the proposed $J$-matrix analysis, a new method that makes it possible to
verify a consistency of  shell model results with experimental phase
shifts.  To the best of our knowledge, this is the only method which can
relate the shell model results to the scattering data in the case of
non-resonant scattering like the $\frac12^+$ $n\alpha$ and $p\alpha$
scattering. In the case of negative parity resonances in $^5$He and
$^5$Li discussed here, the $J$-matrix analysis generally suggests that
the shell model should generate the respective states above the
resonance energies supplemented by their widths. Note that the
$J$-matrix $E_{\lambda=0}$ only in some cases lie inside shaded areas
showing the resonance energies together with their widths
in Figs. \ref{5He} and \ref{5Li}, and in all these cases, the
intersection of the  $E_{\lambda=0}$ with the resonance is seen only at
small enough $\hbar\Omega$ values where the shell model predictions fail
to follow the $J$-matrix analysis results. This is a clear indication
that one should be very accurate in relating the shell model results to
the resonance energies, at least in the case of wide enough resonances.

\section{Conclusions}

We suggest a method of $J$-matrix inverse scattering analysis of elastic
scattering phase shifts and test this method in applications to
$n\alpha$ and $p\alpha$ elastic scattering. We demonstrate that the
method is able to reproduce $\frac32^-$, $\frac12^-$ and $\frac12^+$
$n\alpha$ and $p\alpha$ elastic scattering phase shifts with high
accuracy in a wide range of the parameters of the method like
the oscillator spacing $\hbar\Omega$, model space and the channel radius
$b$ in the case of $p\alpha$ scattering. The method is very simple in
applications, it involves only a numerical solution of a simple
transcendental equation (\ref{Elam}).

When the $J$-matrix phase shift parameterization is obtained, the
resonance parameters, resonance energy and width, can be obtained by
locating the $S$-matrix pole by
solving numerically another simple transcendental equation
(\ref{Spole}). The resonance energies and widths are shown to be
stable when  $\hbar\Omega$ or other $J$-matrix parameters are
varied. Our results for $\frac32^-$ and $\frac12^-$ resonant states in
$^5$He and $^5$Li are compared in Table \ref{tab-comp} with the results
of other authors. Our results are in line with the results of other
studies; in general, the better agreement is seen with
Ref. \cite{Rmatr}, the most recent among all publications presented in
the Table. Cs\'{o}t\'{o} and Hale performed two different analyses in
Ref. \cite{Rmatr}: (i) RGM search for the  $S$-matrix poles based on a
complicated enough calculations within the Resonating Group Model with
effective Minnesota $NN$ interaction fitted to
the
nucleon-$\alpha$ phase shifts, and (ii) Extended $R$-matrix analysis of
$^5$He and $^5$Li including not only $N+\alpha$ channel but also $d+t$
or $d+\rm^3He$ channels along with pseudo-two-body
configurations to represent the breakup channels $n+p+t$ or
$n+p+\rm^3He$ and using a wide range of data on various reactions. We
note that our very simple $J$-matrix approach uses only a very limited
set of data as an input, $n\alpha$ or $p\alpha$ phase shifts.
We suppose that the proposed approach can be useful in analysis of
elastic scattering in other nuclear systems and serve as an alternative
to the conventional $R$-matrix analysis.

\begin{table*}
\caption{Parameters of the low-energy $^5$He and $^5$Li resonances in
 the center-of-mass frame. All numbers are in MeV. {
The uncertainty of  our results was evaluated as the maximal deviation 
 in calculations in model spaces ranging from
 $6\hbar\Omega$ to $14\hbar\Omega$ and with $\hbar\Omega$ values ranging
 from 10 to 30 MeV.}}
\begin{ruledtabular}
\begin{tabular}{l@{\,}l@{\,}l@{\,}l@{\,}l@{\,}l@{\,}l@{\,}l@{\,}l}
  &\multicolumn{4}{c}{$^5$He} &\multicolumn{4}{c}{$^5$Li}\\ \cline{2-5}\cline{6-9}
Method & $E_{res}(3/2^-)$ & $\Gamma(3/2^-)$ & $E_{res}(1/2^-)$ & $\Gamma(1/2^-)$
    & $E_{res}(3/2^-)$ & $\Gamma(3/2^-)$ & $E_{res}(1/2^-)$ & $\Gamma(1/2^-)$\\
                   \hline
Compilation \cite{AjSe} & $0.89\pm 0.05$ & $0.60\pm 0.02$
   & $4.89\pm 1$ & $4\pm 1$ & $1.96.\pm 0.05$ & $\approx 1.5$ &7$-$12 & $5\pm 2$\\
$R$-matrix, stripping \cite{AusJPh} &$0.838\pm 0.018$ & $0.645\pm 0.046$
   &$2.778\pm 0.46$ &$3.6\pm 1.2$ &$1.76\pm 0.06$ &$1.18\pm 0.13$
                    & $3.63\pm 0.56$ & $4.1\pm 2.5$\\
$R$-matrix, pickup \cite{AusJPh} &$0.869\pm 0.003$ &$0.723\pm 0.019$
  &$3.449\pm 0.4$ &$5.3\pm 2.3$ &$1.86\pm 0.01$ &$1.44\pm 0.08$
                   &$4.54\pm 0.5$ &$6.1\pm 2.8$\\
Scattering ampl. \cite{ScatAm} &0.778 &0.639 &1.999 &4.534 &1.637 &1.292
  &2.858 &6.082 \\
$S$-matrix, RGM \cite{Rmatr} &0.76 &0.63 & 1.89 &5.20 &1.67 &1.33 &2.70 &6.25\\
Extended $R$-matrix \cite{Rmatr} &0.80 &0.65 &2.07 &5.57 &1.69 &1.23 &3.18 &6.60\\
$J$-matrix &$0.772\pm 0.005$ &$0.644\pm 0.005$ &$1.97\pm 0.03$ 
  &$5.20\pm 0.05$ & $1.658\pm 0.005$ &
     $1.26\pm 0.01$ & $2.85\pm 0.05$\footnote{
We excluded a single value
obtained in the $14\hbar\Omega$ model space with $\hbar\Omega=10$ MeV in 
our evaluation of this uncertainty (see Fig. \ref{pap-ergamp1}).}&$6.15\pm 0.10$
\end{tabular}
\end{ruledtabular}
\label{tab-comp}
\end{table*}

A very interesting and important output of the $J$-matrix inverse
scattering analysis of the phase shifts is the set of $E_\lambda$ values
which are directly related to the eigenenergies obtained in the shell
model or any other model utilizing the oscillator basis, for example,
the Resonating Group Model. The $J$-matrix parameterizations provide the
energies of the states that should be obtained in the shell model or
Resonating Group Model to generate the given phase shifts. These
energies are shown to be model space and $\hbar\Omega$-dependent and
 very different from the energies of at least
wide enough resonances which are conventionally used to compare with the
shell model results. More, the $J$-matrix analysis is shown to provide
the shell model energies even in the case of non-resonant scattering
such as the $\frac12^+$
nucleon--$\alpha$ scattering.

Our comparison of the lowest $E_{\lambda=0}$ with
the No-core Shell Model
results shows that the shell model fails to
reproduce the phase shifts if  small $\hbar\Omega$
values are employed in the calculations. When $\hbar\Omega$ and/or model
space
size is increased, the shell model predictions approach
$E_{\lambda=0}$ values obtained in the $J$-matrix signaling that the
shell model results become more and more consistent with the experimental
phase shifts. However some difference between the  No-core Shell Model
predictions and the $J$-matrix analysis results is seen
even in the largest model spaces used in this study. This difference is
really not large, its possible sources are the following. (i) There is
an ambiguity in the threshold energies used to relate the absolute
negative energies obtained in the shell model and positive $E_\lambda$
values defined relative to the reaction threshold. (ii) Unfortunately,
there is no $NN$ interaction providing correct energies for, at least,
light nuclei. The JISP16 $NN$ interaction is good enough and provides
reliable predictions for energies of levels in all $s$ and $p$ shell
nuclei \cite{JISP16,JISPYaF,extrap08}. However, there are small differences between
JISP16 level energy predictions and experiment; these differences
are of the same order as the differences
between the $J$-matrix  $E_{\lambda=0}$ values and our  No-core Shell Model
results. Probably we shall use the $J$-matrix results discussed above
while attempting to design a new improved version of the JISP16
interaction by trying to eliminate
the discrepancy between the shell model results and the
$J$-matrix analysis of nucleon-$\alpha$ scattering.

Of course, the $J$-matrix can be used to relate the shell model energies
and data on nucleon scattering by other nuclei. Generally, one can
also use other elastic scattering data, for example, nucleus-nucleus elastic
scattering phase shifts to get the  $E_{\lambda}$ values that should be
obtained in the shell model studies of the respective compound nuclear
systems: the shell model
must generate the states with the same
energies in the same model space and with the same $\hbar\Omega$ value
to have a chance to generate the experimental phase shifts.\\

We are thankful to G. M. Hale
and P. Maris for valuable discussions and help in our
studies.
This work was supported in part by the Russian Foundation of Basic Research, by the US
DOE Grants DE-FC02-07ER41457 and DE-FG02-87ER40371.



\begin{thebibliography}{99}
\bibitem{Lane} A. M. Lane and R. G. Thomas, Rev. Mod. Phys. {\bf 30}, 257
(1958).

\bibitem{YaFi} H. A. Yamani and L. Fishman, J. Math. Phys. {\bf 16}, 410
(1975).

\bibitem{Ztmf1}
S.~A.~Zaytsev, Teoret. Mat. Fiz. {\bf 115}, 263 (1998) [Theor.
Math. Phys. {\bf 115}, 575 (1998)].

\bibitem{ISTP} A. M. Shirokov, A. I. Mazur, S. A. Zaytsev, J. P. Vary,
    and T. A. Weber, Phys. Rev. C {\bf 70}, 044005 (2004);  
    {
 in {\itshape The $J$-Matrix  Method. Developments and Applications}, 
        edited by A. D. Alhaidari, H. A. Yamani, E. J. Heller, and
        M. S. Abdelmonem (Springer, 2008), 219}.

\bibitem{JISP6} A. M. Shirokov, J. P. Vary, A. I. Mazur, S. A. Zaytsev,
    and T. A. Weber, Phys. Lett. {\bf B621}, 96 (2005); J. Phys. G
    {\bf 31}, S1283 (2005).

\bibitem{JISP16} A. M. Shirokov, J. P. Vary, A. I. Mazur, and
    T. A. Weber, Phys. Lett. {\bf B644}, 33 (2007).

\bibitem{Bang} J. M. Bang, A. I. Mazur, A. M. Shirokov, Yu. F. Smirnov,
and S. A. Zaytsev, Ann. Phys. (NY) {\bf 280}, 299 (2000).

\bibitem{Vary} D.~C. Zheng, J. P. Vary, and B.~R.~Barrett, Phys. Rev. C
{\bf 50}, 2841 (1994); D.~C.~Zheng, J.~P.~Vary,  B.~R.~Barrett,
W.~C.~Haxton, and C.~L.~Song, Phys. Rev. C {\bf 52}, 2488 (1995).


{
\bibitem{SmSh} A. M. Shirokov, Yu. F. Smirnov, and S. A. Zaytsev,
in {\itshape Modern Problems in Quantum Theory}, edited by V.~I.~Savrin and
O.~A.~Khrustalev, (Moscow State University, Moscow, 1998), 184;
Teoret. Mat. Fiz. {\bf 117}, 227 (1998) 
[Theor. Math. Phys. {\bf 117}, 1291 (1998)].

\bibitem{Fil} G. F. Filippov and I. P. Okhrimenko, Yad. Fiz. {\bf 32}, 932
(1980) [Sov. J. Nucl. Phys. {\bf 32}, 480 (1980)]; G.~F.~Filippov, Yad.
Fiz. {\bf 33}, 928 (1981) [Sov. J. Nucl. Phys. {\bf 33}, 488 (1981)].  }

\bibitem{Kuku} V. I. Kukulin, V. N. Pomerantsev, Kh. D. Razikov,
V. T. Voronchev, and G. G. Ryzhikh, Nucl. Phys. A {\bf 586}, 151 (1995).

{
\bibitem{Okhr} I. P. Okhrimenko,  Nucl. Phys. A {\bf 424}, 121 (1984).
}

\bibitem{ArndtDLRop} R. A. Arndt, D. D. Long, and L. D. Roper,
    Nucl. Phys. A {\bf 209}, 429  (1973).


\bibitem{NPA287} J. E. Bond and F. W. K. Firk, Nucl. Phys. A {\bf 287},
    317 (1977).

\bibitem{Rmatr} A. Cs\'{o}t\'{o} and G. M. Hale, Phys. Rev. C {\bf
    55},536 (1997).


\bibitem{Rep-na} B. V. Danilin, M. V. Zhukov, A.~A.~Korsheninnikov, and
L.~V.~Chulkov, Yad. Fiz. {\bf 53}, 71 (1991) [Sov. J. Nucl. Phys. {\bf 53}, 45
(1991)].

\bibitem{Forb-na} J.~Bang and C.~Gignoux, Nucl. Phys. {\bf A~313}, 119 (1979).

\bibitem{Forb-na2}  Yu.~A.~Lurie and A.~M.~Shirokov,  Izv. Ros. Akad. Nauk,
Ser. Fiz. {\bf 61}, 2121 (1997) [Bull.
Rus. Acad. Sci., Phys. Ser. {\bf 61}, 1665 (1997)].

\bibitem{LurAnn} Yu.~A.~Lurie and A.~M.~Shirokov,
Ann. Phys. (NY) {\bf 312}, 284 (2004); 
        {
in {\itshape The $J$-Matrix Method. Developments and
        Applications}, edited by A. D. Alhaidari, H. A. Yamani, E. J. Heller, and
        M. S. Abdelmonem (Springer, 2008), 183}.

\bibitem{IS} The forbidden state in this model is a particular case of
    so-called isolated states, see
A.~M.~Shirokov and S.~A.~Zaytsev, in {\itshape The $J$-Matrix
        Method. Developments and
        Applications}, edited by A. D. Alhaidari, H. A. Yamani, E. J. Heller, and
        M. S. Abdelmonem (Springer, 2008), 103; quant-ph/0312065 (2003);
S.~A.~Zaytsev,  Yu.~F.~Smirnov, and A.~M.~Shirokov, Izv. Ros. Akad. Nauk,
Ser. Fiz. {\bf 56}, 80 (1992).


\bibitem{Arndt} R. A. Arndt, L. D. Roper, and R. L. Shotwell, Phys. Rev.
C {\bf 3}, 2100 (1971).

\bibitem{NPA163} P. Schwandt, T.~B.~Clegg, and
    W.~Haeberli. Nucl. Phys. A {\bf 163}, 432 (1971).

\bibitem{Dodder} D. C. Dodder, G. M. Hale, N.~Jarmie, J.~H.~Jett,
P.~W.~Keaton, Jr., R.~A.~Nisley, and K.~Witte. Phys. Rev. C {\bf 15},
    518 (1977).


\bibitem{Vary92_MFDn} J. P. Vary, ``The
Many-Fermion-Dynamics Shell-Model Code," Iowa State University,
1992 (unpublished); J. P. Vary and D. C. Zheng, {\it ibid} 1994  (unpublished);
test runs can be performed through http://nuclear.physics.iastate.edu/mfd.php.

\bibitem{JWEB} FORTRAN code generating JISP16 interaction is available at
http://nuclear.physics.iastate.edu/.

\bibitem{AjSe} F. Ajzenberg-Selove, Nucl. Phys. A {\bf 490}, 1 (1988).

\bibitem{AusJPh} C. L. Woods, F. C. Barker, W. N. Catford,
    L. K. Fifield, and N. A. Orr, Aust. J. Phys. {\bf 41}, 525,
    (1988);   F. C. Barker and C. L. Woods, {\em ibid.} {\bf 38},
    563 (1985).

\bibitem{ScatAm}   M. U. Ahmed and P. E. Shanley, Phys. Rev. Lett.
{\bf 36}, 25 (1976).

\bibitem{JISPYaF} A. M. Shirokov, J. P. Vary, A. I. Mazur, and
    T. A. Weber, Yad. Fiz.
{\bf 71}, 1260 (2008) [Phys. At. Nucl., {\bf 71}, 1232 (2008)].

\bibitem{extrap08} P. Maris, J. P. Vary, and A. M. Shirokov, 
{
arXiv:0808.3420 (2008)}.

\bibitem{Roth} R. Roth, J. R. Gour and P. Piecuch, arXiv:0806.0333
	(2008).

\end{thebibliography}
\end{document}